\pdfoutput=1 

\documentclass[sigconf]{acmart}

\usepackage{graphicx}
\usepackage{balance}  %
\usepackage{algorithm}
\usepackage{algpseudocode}
\usepackage{fix-cm}
\usepackage{multirow}
\usepackage{caption}
\usepackage{subcaption}
\usepackage{amsmath}
\usepackage{bbold}
\usepackage{mathtools}
\usepackage{xcolor}
\usepackage{tablefootnote}
\usepackage{threeparttable}
\usepackage{hyperref}
\usepackage{amsthm}
\usepackage{enumitem}
\usepackage{pifont}
\usepackage{gensymb}
\usepackage{listings}
\usepackage{xr} 
\usepackage{wrapfig}
\usepackage{setspace}
\usepackage{tablefootnote}
\usepackage[colorinlistoftodos,prependcaption,textsize=tiny]{todonotes}
\usepackage{varwidth}
\newcommand{\META}[1]{}
\newcommand{\METACAP}[1]{}
\newcommand{\RI}[1]{}
\newcommand{\RICAP}[1]{}
\newcommand{\RII}[1]{}
\newcommand{\RIICAP}[1]{}
\newcommand{\RIII}[1]{}
\newcommand{\RIIICAP}[1]{}

\newcommand{\stitle}[1]{\noindent \textbf{#1} }

\newcommand{\thr}{{\tt thres}}
\newcommand{\rhat}{\hat{R}}

\newcommand{\phat}{\hat{p}}

\newcommand{\tf}{{\mathbb{k}_i}}
\newcommand{\pf}{{\mathbb{p}_i}}
\newcommand{\tfj}{{\mathbb{k}_j}}
\newcommand{\pfj}{{\mathbb{p}_j}}
\newtheorem*{definition}{DEFINITION}

\newcommand{\cmark}{\ding{51}}%
\newcommand{\xmark}{\ding{55}}%

\newcommand{\name}{\textsc{Everest}}

\DeclarePairedDelimiter\floor{\lfloor}{\rfloor}

\algnewcommand\algorithmicforeach{\textbf{for each}}
\algdef{S}[FOR]{ForEach}[1]{\algorithmicforeach\ #1\ \algorithmicdo}
\algnewcommand\And{\textbf{and }}
\algnewcommand\Or{\textbf{or }}

\usepackage{xcolor}  %
\definecolor{Plum}{HTML}{C2938D}
\definecolor{light-gray}{gray}{0.8}
\newcommand{\deleted}[1]{{\color{light-gray}#1}}
\newcommand{\added}[1]{{#1}}

\newcommand{\replaced}[2]{\deleted{#1}\added{#2}}

\renewcommand{\deleted}[1]{}

\definecolor{deepblue}{rgb}{0,0,0.5}
\definecolor{deepred}{rgb}{0.6,0,0}
\definecolor{deepgreen}{rgb}{0,0.5,0}

\lstdefinestyle{mystyle}{
    commentstyle=\color{codegreen},
    keywordstyle=\bf\color{deepblue},
    numberstyle=\tiny\color{black},
    stringstyle=\color{deepgreen},
    basicstyle=\ttfamily\footnotesize,
    breakatwhitespace=false,         
    breaklines=true,                 
    captionpos=b,                    
    keepspaces=true,                 
    showspaces=false,                
    showstringspaces=false,
    showtabs=false,                  
    tabsize=2
}
\usepackage{array}
\newcolumntype{M}[1]{>{\centering\arraybackslash}m{#1}}
\newcolumntype{P}[1]{>{\centering\arraybackslash}p{#1}}

\copyrightyear{2021}
\acmYear{2021}
\setcopyright{acmcopyright}\acmConference[SIGMOD '21]{Proceedings of the 2021 International Conference on Management of Data}{June 18--27, 2021}{Virtual Event , China}
\acmBooktitle{Proceedings of the 2021 International Conference on Management of Data (SIGMOD '21), June 20--25, 2021, Virtual Event , China}
\acmPrice{15.00}
\acmDOI{10.1145/3448016.3452786}
\acmISBN{978-1-4503-8343-1/21/06}

\renewcommand\footnotetextcopyrightpermission[1]{} %
\begin{document}
\title{Top-K Deep Video Analytics: A Probabilistic Approach}
\setcounter{page}{1}

\author{Ziliang Lai$^*$, Chenxia Han$^*$, Chris Liu$^*$, Pengfei Zhang$^*$, Eric Lo$^*$, Ben Kao$^\dagger$}

\affiliation{\institution{$^*$The Chinese University of Hong Kong}\country{}}
\email{{zllai,cxhan,cyliu,pfzhang,ericlo}@cse.cuhk.edu.hk}
\affiliation{\institution{$^\dagger$University of Hong Kong}\country{}}
\email{kao@cs.hku.hk}


\begin{abstract}
The impressive accuracy of deep neural networks (DNNs) has created great demands on practical analytics over video data.   Although efficient and accurate, the latest video analytic systems have not supported analytics beyond selection and aggregation queries.  In data analytics, Top-K is a very important analytical operation that enables analysts to focus on the most important entities.  In this paper, we present \name, the first system that supports efficient and accurate Top-K video analytics.  
\name~ ranks and identifies the most interesting frames/moments from videos with probabilistic guarantees.  
\name~is a system built with a careful synthesis of deep computer vision models, uncertain data management, and Top-K query processing.
Evaluations on real-world videos and the latest Visual Road benchmark show that \name~ achieves between \added{14.3$\times$ to 20.6$\times$} higher efficiency than baseline approaches with high result accuracy.

\end{abstract}
\settopmatter{printfolios=true}

\fancyhead{}
\pagenumbering{gobble}
\maketitle

\section{Introduction} \label{sec:introduction}
Cameras are ubiquitous. Billions of them are being deployed 
in public (e.g., at road junctions)
and private (e.g., in retail stores) all over the world \cite{cctv}. 
Recent advances in deep convolutional neural networks (CNNs) have led to 
incredible leaps in their accuracies in many machine learning tasks, notably image and video analysis. 
Such developments have created great demands on practical analytics over video data
\cite{miris, blazeit, noscope, scanner, videostorm}.

\replaced{Object recognition in image and video using CNNs are accurate \cite{he2017mask, resnet, prelu}}
{Deep CNNs can extract rich semantic 
(e.g., object detection \cite{he2017mask}, facial sentiment \cite{sentiment}, and 3D-depth estimation \cite{depth_estimation})
from videos}.
Nonetheless, CNNs are also computationally expensive to train and to serve. 
For example, 
state-of-the-art object detectors
run at about 5 fps
(frames per second) using GPUs \cite{he2017mask}. This frame-processing rate is 6 times slower than the frame rate of a typical 30-fps video. 
Consequently, 
a naive ``scan-and-test'' approach that invokes, say, an object detector on every frame
would take 6 times the duration of a video to complete an object-related query.
While we can parallelize that by, say, using multiple GPUs, 
the computational expense remains high regardless of parallelism.
Hence, the research community has started to build systems with innovative solutions to support fast analytics over video data
\cite{tahoma, miris, rekall, focus, chameleon, noscope, blazeit, optasia, pp, vista, scanner, SVQ, videostorm, panorama}.

Deep video analytics is an emerging research topic that intersects database and computer vision.
It is, however, still in its infancy.
For example, latest systems 
support only \emph{selection} queries 
such as object selection \cite{blazeit, noscope, focus, tahoma, pp, exsample, vstore, tasm} 
and 
object-trajectory selection \cite{miris, deeplens}.
In data analytics, Top-K is a very important analytical operation 
that enables analysts to focus on the most important 
entities in the data \cite{saying-enough, survey-topk, bruno, top-k-guarantee}.  
In this paper, we present
the very first system for efficient Top-K video analytics.  
Top-K can help rank and identify the most interesting frames/moments from videos.  
Example use cases include:

\stitle{Property Valuation.}  
The valuation/rent of a shop is strongly related to its peak foot traffic \cite{foot-traffic}.
Instead of manual counting, one can use a camera to capture the pedestrian flow and use a deep object detector to identify, say, the Top-5 frames (time of the day) with the \emph{highest pedestrian counts}.

\deleted{
\stitle{Data-Driven Agriculture.}
The global food demand is expected to increase by 70\% from
2010 to 2050
\cite{food-demand}.
In order to improve farm productivity, 
the FarmBeats project at Microsoft sends drones to collect and analyze farm videos \cite{farmbeats}.
Recent news also report that oil-palm farmers in Malaysia send drones to monitor the growth of oil-palm \cite{raghu_2019}. 
With oil-palm plantations spread across 86,100 square miles in Malaysia,
farmers with limited resources can only inspect a small number of fields onsite each day.
Finding the Top-K fields (e.g., based on number of well-grown palm trees) over drone videos can drastically help farmers prioritize field trips.
}

\added{
\stitle{Thumbnail Generation.} \RI{R1O1}\RII{R2O1O2}
Thumbnails are images of videos 
that could largely affect the click-through rate of a video \cite{click-through}. 
To select attractive thumbnails, 
a social platform can use a deep visual sentimentalizer \cite{sentiment}
to extract, say, the Top-10 \emph{happiest moments} of a video as the thumbnails for grabbing the viewers' attention. 
}

\deleted{
\stitle{Transport Planing.}  
Severe traffic congestion can occur when multiple bus routes pass through the same narrow expressway.
A Top-K query on the moments with the biggest bus convoy
provides valuable information on the congestion caused
and for better bus scheduling.
}

\added{
\stitle{Fleet Management.}
Fleet managers in trucking businesses 
analyze dashcam videos 
to flag dangerous driving behaviors of their truck drivers \cite{allen_2020}. 
With a deep depth estimator \cite{depth_estimation},
a fleet manager can query, say, the Top-50 dashcam
frames based on the distance between 
the truck and its front vehicle 
as the most \emph{dangerous tailgating moments}
to assess drivers' safety awareness.
}

In video analytics, \emph{CNN Specialization}
and \emph{Approximate Query Processing} (AQP)
are two popular techniques 
for overcoming the inefficiency of the  scan-and-test method \cite{noscope, tahoma, focus, blazeit}.
With CNN specialization, 
a video analytic system trains a 
lightweight \emph{query-specific model}
(e.g., a model specifically for ``bus'' detection)
using frames sampled from the video-of-interest as the training data.
Since the specialized CNN is highly specific to the query instance and the video-of-interest,
it is both efficient to train and to serve (but is generally less accurate).
The specialized CNN is then applied to 
all video frames to obtain a set of candidate frames that are likely to satisfy the given query.
Finally, the candidate frames are passed back to a
\added{general deep model that serves as an accurate but slow-to-run ``oracle'' to verify their validity}. 
Approximate Query Processing (AQP) \cite{blazeit, BlinkDB} is 
an orthogonal 
technique for fast analytics 
when users are willing to tolerate some statistical error.
The complex interplay between deep model inference and Top-K query processing,
however, creates novel challenges for CNN specialization and AQP.

First, unlike selection queries where a frame satisfies a predicate or not is \emph{independent} of other frames,
whether a frame is in Top-K 
requires \emph{score comparisons between it and others}. 
Hence,
short-circuiting 
selections on \emph{individual} frames 
using
specialized CNNs (e.g., \cite{noscope,focus,tahoma}) 
is insufficient to Top-K query processing that requires score comparisons.
Second, AQP techniques 
for video analytics (e.g., \cite{blazeit})
estimate \emph{statistics} (e.g., average number of cars) 
but Top-K query returns a \emph{set} (of frames) instead of a statistic,
which demands another notion of approximation for Top-K under the video setting.
Third, video data has temporal locality
and thus, Top-K queries can be frame-based or 
\emph{window-based}. 
For example, one could be interested in finding the Top-K 5-second clips with the
highest number of vehicles.
This adds to the complexity of answering Top-K queries in video analytics.

To address these challenges,
we present \name, a system that empowers users to pose Top-K
and Top-K window queries
on videos based on any ranking function \added{
defined over the scores provided by deep models (e.g., sentiment scores)}. \RI{R1O1}\RII{R2O1O2}
Since \emph{model inference results are 
intrinsically probabilistic}, 
\name~treats {\bf inference score distribution as a first-class citizen} 
so that it can offer {\bf probabilistic guarantees} (e.g., guaranteeing a Top-K result has 99.99\% chance of being the exact result).
This design is in sharp contrast to existing video analytic systems where valuable uncertainty information is largely discarded.
To illustrate, Table \ref{tab:sota} shows an example output of the specialized CNN used in BlazeIt \cite{blazeit}.
The (car) \texttt{count} in each traffic video frame is best given in the form a \emph{probability distribution} by the \texttt{softmax} layer of the classifier, which captures
the \emph{uncertainty} of the prediction.
Those systems, however, process queries based on 
a \emph{trimmed} view where 
\emph{all cases except the most probable one are discarded}
(see Table \ref{tab:blazeit}).

Supporting Top-K analytics over videos 
enables rich analyses {over the visual world}, 
just as what traditional Top-K query processing has done over relational data.
Beyond system design, \name~has to develop new uncertain Top-K query processing algorithms 
based on its {novel setting}.
\replaced{

(e.g., 
they mostly focus on 
the human efficiency issues 
(e.g., 
the plurality of a task that is needed to improve answer accuracy, given that human workers make mistakes
\cite{davidson2013using,  kou2017crowdsourced}). 
By contrast, \name~focuses on 
the system and algorithm designs
to perform
\textbf{probabilistic Top-K video analytics at scale}.
}
{
Specifically, 
existing Top-K uncertain query processing techniques 
that are most related to \name~are those 
that use human oracles to reduce uncertainties online
while evaluating a Top-K query \cite{zhang2015cleaning, chu2015katara}.
Nonetheless, they 
offer no probabilistic guarantees and focus on {human efficiency} aspects instead, 
e.g., the plurality of a task that is needed to improve answer accuracy with considerations of human mistakes and budget limitation \cite{davidson2013using,  kou2017crowdsourced}.
\name, in contrast, focuses on the scalability, probabilistic guarantees, and windowing issues.

}

\replaced{
\begin{enumerate}
    \item We design and present a novel notion of Top-K analytics over videos.
    Our design is the first to treat 
    the probability distribution from specialized CNNs
    as a first-class citizen in video analytics.

    \item We present \name, 
    the first video analytics system that supports Top-K queries with probabilistic guarantees.
    The system is able to support  
    both frame-based and window-based queries
    and support different types of ranking functions.

    \item We develop efficient
    algorithms
    and implementations for each system module.  Our algorithms overcome the combinatorial explosion 
in the number of possible worlds commonly found in
    uncertain query processing.
    Our system is highly optimized using techniques 
    such as batch inference and prefetching.

\end{enumerate}
}
{
To summarize, the paper makes the following principled contributions. \RII{R2O1}
System-wise, \name~is the first deep video analytics system to 
(i) serve Top-K queries with deep semantic at scale;
(ii) connect the worlds of uncertain data management with deep visual models; and 
(iii) provide probabilistic query guarantees.
Algorithm-wise, \name~addresses the first uncertain Top-K problem 
with the presence of accurate but slow-to-run machine learning oracles.
We have evaluated \name~using different 
analytical tasks on different real-world videos as well as 
using the latest Visual Road benchmark \cite{visualroad}.
Experimental results show that \name~can achieve \added{14.3$\times$ to 20.6$\times$} speedup over different baseline approaches with high accuracy.

}

The remainder of this paper is organized as follows.
Section \ref{sec:background_related_work} provides essential background \added{and related works} of this paper.
Section \ref{sec:Topk} discusses \name~in detail.
Section \ref{sec:exp} gives the evaluation results.
\deleted{Section \ref{sec:related} discusses related work.}
Finally, Section \ref{sec:conclusion} concludes the paper.

\begin{table}\small
    \begin{subtable}{0.54\linewidth}
    \centering
        \begin{tabular}[b]{|c|c|c|}
            \hline
            \begin{tabular}[c]{@{}c@{}}{\sf timestamp}\\ {\sf /frame}\end{tabular} & {\sf count} & {\sf prob.} \\
            \hline\hline
            \multirow{3}{*}{$f_1$} & \multicolumn{1}{c|}{0} & \multicolumn{1}{c|}{0.78} \\\cline{2-3}
                                   & \multicolumn{1}{c|}{1} & \multicolumn{1}{c|}{0.21} \\\cline{2-3}
                                   & \multicolumn{1}{c|}{2} & \multicolumn{1}{c|}{0.01} \\\hline
            \multirow{3}{*}{$f_2$} & \multicolumn{1}{c|}{0} & \multicolumn{1}{c|}{0.49} \\\cline{2-3}
                                   & \multicolumn{1}{c|}{1} & \multicolumn{1}{c|}{0.42} \\\cline{2-3}
                                   & \multicolumn{1}{c|}{2} & \multicolumn{1}{c|}{0.09} \\\hline
            \multirow{3}{*}{$f_3$} & \multicolumn{1}{c|}{0} & \multicolumn{1}{c|}{0.16} \\\cline{2-3}
                                   & \multicolumn{1}{c|}{1} & \multicolumn{1}{c|}{0.48} \\\cline{2-3}
                                   & \multicolumn{1}{c|}{2} & \multicolumn{1}{c|}{0.36} \\\hline
        \end{tabular}
        \caption{Output of a lightweight model}
        \label{tab:sota}
    \end{subtable}
    \hfill
    \begin{subtable}{0.45\linewidth}
    \centering
        \begin{tabular}[b]{|c|c|}
            \hline
            \begin{tabular}[c]{@{}c@{}}{\sf timestamp}\\ {\sf /frame}\end{tabular} & {\sf count} \\
            \hline\hline
              $f_1$ & \multicolumn{1}{c|}{0}  \\\hline
            $f_2$ & \multicolumn{1}{c|}{0}  \\\hline
            $f_3$ & \multicolumn{1}{c|}{1}  \\\hline
        \end{tabular}
        \caption{Existing systems discard uncertainty information}
        \label{tab:blazeit}
    \end{subtable}
\caption{State-of-the-art}
\end{table}

\replaced{\input{background}}{\section{Background and Related Work}\label{sec:background_related_work} \RIII{R4O2O3}
\name~builds upon research in modern video analytics and uncertain data management.
In this section, we provide 
essential background as well as a discussion on the related works.

\added{
\stitle{Traditional Visual Data Management Systems.} \RII{R2O2}
Pioneered by Chabot \cite{chabot} and QBIC \cite{qbic}, traditional visual data management systems feature ``content-based'' retrieval that takes as input an example image, a sketch or a textural description of rough appearance, and outputs similar images or video segments \cite{sbhf, region-based, jacob, strg_index, browsing_indexing, content_based_survey, mpeg_method}. 
However, many of these systems are based on matching low-level features (e.g., colors). To query richer semantic, additional human annotations are often required \cite{shiatsu, vdbms}. 
Top-K queries in these systems mostly 
leverage the geometric properties of the similarities to early-stop the process \cite{qbic, shiatsu}. In contrast, \name~focuses on Top-K queries that rank frames based on scores from deep models (e.g., sentiment scores), where no such geometric properties are available.
}

\stitle{Deep Video Analytics Systems.}
With the advent of convolutional neural networks (CNNs), 
richer \textit{semantic} of the video can be extracted by machines with accuracy close to or even surpass human \cite{prelu, geirhos2017comparing}. 
For instance, object detectors can identify objects and their  
bounding boxes in an image \cite{he2017mask, yolo3}; 
visual sentimentalizers 
can evaluate the degree of happiness or sadness in an image \cite{sentiment}; 
and depth estimators 
can estimate 3D-depth from an image \cite{depth_estimation}. 
Nonetheless, with millions of parameters in these deep CNNs,
their high inference costs pose great challenges to 
modern video analytics research \cite{noscope,focus,scanner,tahoma, videostorm, blazeit, exsample}.

For analytical processing, 
video data is often modeled as relations, which captures the extracted semantics of the video. Table \ref{tab:data} shows an example of the video relation model used in BlazeIt \cite{blazeit}, which is populated by an object detector.
Specifically,
each tuple in the relation corresponds to a single object in a video frame. 
Since a frame may contain 0 or more objects (of interest), and an object may appear in multiple
frames, a frame can be associated with 0 or more tuples in a relation and an object can be associated with
multiple tuples. 
Typical attributes of a tuple include
a frame timestamp 
({\sf ts}),
a unique id of an identified object ({\sf objectID}),
the object's class label ({\sf class}), 
 bounding polygon ({\sf polygon}),
 raw pixel content ({\sf content}),
and feature vector ({\sf features}).
To recognize identical objects across frames so that they share the same {\sf objectID},  
an object tracker is invoked (e.g, \cite{zhang2011video}), 
which takes as input two polygons 
from two consecutive frames and returns the same {\sf objectID} if the two polygons 
represent the same object.
In video analytics, a video relation that is materialized by an accurate deep CNN such as YOLOv3 \cite{yolo3}
is regarded as the \textit{ground-truth} \cite{focus}.
However, fully materializing a ground-truth relation is computationally expensive.
Therefore, 
the key challenge 
is how to answer queries without 
fully materializing that video relation 
\cite{noscope,focus,scanner,tahoma,videostorm,blazeit,exsample}.

\begin{table}
    \centering
    \resizebox{\linewidth}{!}{
    \begin{tabular}{c|c|l|c|c|c}
{\sf Timestamp (ts)} &   {\sf Class}  & {\sf Polygon} & {\sf ObjectID} & {\sf Content} & {\sf Features}   \\\hline\hline
01-01-2019:23:05 &   Human & (10, 50), (30, 40), $\dots$ & 16 &  $\dots$  &$\dots$   \\
01-01-2019:23:05 &   Bus   & (45, 58), (66, 99), $\dots$ & 58 &  $\dots$  &$\dots$   \\\hline 
01-01-2019:23:06 &   Human & (20, 80), (7, 55), $\dots$ & 16 &  $\dots$  &$\dots$   \\
01-01-2019:23:06 &   Car   & (6, 91), (10, 55), $\dots$ & 59 &  $\dots$  &$\dots$   \\
01-01-2019:23:06 &   Car   & (78, 91), (40, 55), $\dots$ & 60 &  $\dots$  &$\dots$   \\\hline 
$\dots$ & $\dots$ & $\dots$ & $\dots$ & $\dots$ & $\dots$  \\

\end{tabular}
}
\caption{A video relation fully populated by a ground-truth object detector}
\vspace{-0.5cm}
\label{tab:data}
\end{table}

\begin{table}
    \centering
    
    \resizebox{\linewidth}{!}{
    
    \begin{tabular}{|m{3.5cm}|P{2cm}|c|c|} \hline
\multicolumn{1}{|c|}{\sf System } &   {\sf Query type}  & \begin{tabular}[c]{@{}c@{}}{\sf Probabilistic} \\ {\sf guarantee}\end{tabular} & \begin{tabular}[c]{@{}c@{}}{\sf Window} \\ {\sf support}\end{tabular}   \\\hline\hline
NoScope \cite{noscope}, Probabilistic predicate \cite{pp}, SVQ \cite{SVQ}, Focus \cite{focus}, ExSample \cite{exsample}, TAHOMA \cite{tahoma} & selection & \xmark & \xmark  \\\hline
\multirow{2}{*}{BlazeIt \cite{blazeit}} & selection & \xmark & \xmark \\\cline{2-4}
                                        & aggregation & \cmark & \xmark \\\hline
MIRIS \cite{miris} & object tracking & \xmark & \cmark \\\hline
\textbf{\name~} & \textbf{Top-K} & \textbf{\cmark} & \textbf{\cmark} \\\hline 
\end{tabular}
}
\caption{\added{Features comparisons among \name~and related deep video analytics systems}}
\vspace{-0.4cm}
\RIIICAP{R4O4}
\vspace{-0.5cm}
\label{tab:related_work}
\end{table}

Table \ref{tab:related_work} summarizes existing video analytic systems that are most related to \name.
NoScope \cite{noscope} supports object selection queries 
based on CNN specialization 
and has an optimizer to select the best model architecture (e.g., number of layers)
that maximizes the throughput subject to a specified accuracy target on a validation set (but no guarantee on the whole video).
BlazeIt \cite{blazeit} develops a SQL-like language for declarative video analytics, but it has no Top-K semantic. BlazeIt optimizes selection similar to NoScope and it also supports aggregation queries, where AQP techniques are used to bound the error. Statistical AQP techniques are inapplicable to \name~because Top-K queries returns a set (of frames) instead of a statistic.
SVQ \cite{SVQ} is similar to BlazeIt with additional support to video streams and spatial constraints.
``Probabilistic predicates'' \cite{pp} can be viewed as a generalization of CNN specialization that allow users to express selection criteria with boolean expressions.
Focus \cite{focus} accelerates selection by pre-building indexes for objects in videos.
ExSample \cite{exsample} is a sampling method to select distinct objects from videos.
TAHOMA \cite{tahoma} speed up object selection by transforming the input images (e.g., reducing resolution).
MIRIS \cite{miris} is a system that supports 
object tracking (i.e., predicates that span across multiple frames). It guarantees accuracy on a validation set but not the whole video.
None of the above systems address Top-K queries.
\name~can guarantee accuracy on the whole video.
Only \name~and MIRIS can support multi-frame analytics through windowing.

\stitle{Uncertain Databases.} \RII{R2O3}
In order to yield high-quality results, 
\name~regards deep models' probabilistic inference results as  first-class citizens.
One common uncertain data representation 
is ``\emph{x-tuples}'' \cite{aggarwal2009trio}.
An \emph{uncertain relation} %
is a collection of x-tuples,
each consists of a number of alternative outcomes that are associated with their corresponding probabilities (e.g., Table \ref{tab:sota}). 
Together, the alternatives form a discrete probability distribution of the true outcome.
\name~avoids materializing the ground-truth video relation by populating an \textit{uncertain relation} using a cheap 
\emph{proxy} to the expensive oracle 
that we will introduce in Section \ref{sec:phase_1}.
x-tuples are assumed to be independent of each other. 
We will discuss later how \name~uses a difference detector
so that the outputs of the proxy scorer
can be represented by the x-tuple model
(i.e., the inference result of a video frame is captured by one x-tuple).
Hence, in the following discussion, 
we use the terms \emph{x-tuple}, 
\emph{frame}, and \emph{timestamp} interchangeably. 

\stitle{Uncertain Top-K Processing.} \RII{R2O3}
There are different notions of uncertain Top-K queries \cite{cormode2009semantics,TIOpaper,soliman2008probabilistic,prob_threshold}.
The notion of \emph{U-TopK} \cite{TIOpaper, soliman2008probabilistic} 
returns a result set that has the highest probability of being Top-K. While U-TopK may return an answer of very low probability (e.g., $10^{-6}$), \name~guarantees the answer meets a probability threshold.
\emph{U-KRanks} \cite{uncertain_db_1, soliman2008probabilistic} is another notion.
In a \emph{U-KRanks} result set, 
the $i$-th result in the result set is the most probable one to be ranked $i$-th. 
However, that does not guarantee that the result set as a whole is the most probable Top-K answer.
\textit{Probabilistic threshold Top-K }\cite{prob_threshold} is yet another notion.  A Top-K result of such kind 
consists of all the tuples (can be less/more than $K$ tuples) 
whose individual probability of being {one of} the Top-K tuples is larger than a given threshold.
There is no 
guarantee that the result set as a whole is the most probable Top-K answer.
For example, it may return an empty set when no tuple satisfies the threshold requirement.
The above uncertain Top-K query processing works have been 
assuming no ground-truth is accessible at run-time. 
In contrast, \name~can access an 
accurate but slow-to-run oracle
to reduce data uncertainties online 
and thus achieve \emph{oracle-in-the-loop uncertain Top-K query processing.}

\stitle{Uncertain Top-K Processing with an Oracle.} \RII{R2O3}
With a slow-to-run but accurate oracle accessible while processing uncertain Top-K queries,  
\name~is related to the topic of 
\emph{uncertain data cleaning} \cite{clean-topk, cheng2008cleaning, zhang2015cleaning}.
There are two branches in uncertain data cleaning.  
The main branch is to identify the best set of uncertain tuples 
for a ``cleaning agent'' (e.g., a human expert) to clean data \emph{offline}.
A human expert is served as an oracle.
As employing a human expert involves monetary cost,
existing works focus on minimizing the expected entropy over the set of possible query answers within a cleaning budget (e.g., 
the number of uncertain tuples to be sent to the cleaning agent) \cite{clean-topk, cheng2008cleaning}. 
In that branch of work, 
the cleaning agent is technically out of the processing loop and an algorithm simply identifies the batch of uncertain tuples that have to be cleaned and terminates. 
In contrast, \name~puts the oracle in the loop and carries out ``cleaning'' \emph{online}.
Hence, it can provide probabilistic guarantees on the query answer, which is way more intuitive than using entropy. 
The other branch of work is closer to \name~because it has an online oracle-in-the-loop setting.
Specifically, that direction leverages 
a crowdsourcing platform (e.g., Amazon Mechanical Turk) 
to reduce the uncertainties online 
while processing an uncertain Top-K query \cite{zhang2015cleaning}.
However, existing works focus more on 
the human efficiency issues.
In contrast, \name~focuses on the scalability, probabilistic guarantees, 
and windowing issues.

}

\begin{table} \small
    \centering
    \subfloat[\tiny{$Pr(W_1) = 0.78 \times 0.49 \times 0.16$}]{
    \begin{tabular}{|c|c|}
      \hline
      {\sf timestamp} & {\sf num of cars}\\
      \hline\hline
      $f_1$ & 0\\
      \hline
      $f_2$ & 0\\
      \hline
      $f_3$ & 0\\
      \hline
    \end{tabular}
    } 
    \subfloat[\tiny{$Pr(W_2) = 0.21 \times 0.49 \times 0.16$}]{
    \begin{tabular}{|c|c|}
      \hline
      {\sf timestamp} & {\sf num of cars}\\
      \hline\hline
      $f_1$ & 1\\
      \hline
      $f_2$ & 0\\
      \hline
      $f_3$ & 0\\
      \hline
    \end{tabular}
    }
    \caption{Two possible worlds $W_1$ and $W_2$}
    \label{tab:possible_world}
\end{table}

\begin{table} \small
    \centering
    \begin{tabular}[b]{|c|c|c|}
    \hline
    {\sf timestamp} & {\sf num of cars} & {\sf conf.} \\
    \hline\hline
      \multirow{3}{*}{$f_1$} & \multicolumn{1}{c|}{0} & \multicolumn{1}{c|}{0.78} \\\cline{2-3}
                           & \multicolumn{1}{c|}{1} & \multicolumn{1}{c|}{0.21} \\\cline{2-3}
                           & \multicolumn{1}{c|}{2} & \multicolumn{1}{c|}{0.01} \\\hline
    \multirow{3}{*}{$f_2$} & \multicolumn{1}{c|}{0} & \multicolumn{1}{c|}{0.49} \\\cline{2-3}
                           & \multicolumn{1}{c|}{1} & \multicolumn{1}{c|}{0.42} \\\cline{2-3}
                           & \multicolumn{1}{c|}{2} & \multicolumn{1}{c|}{0.09} \\\hline
    \multirow{1}{*}{$f_3$} & \multicolumn{1}{c|}{0} & \multicolumn{1}{c|}{1.0} \\\hline
    \end{tabular}
    \caption{After applying \replaced{$GTOD(f_3)$}{$\textsf{Oracle}(f_3)$} on Table \ref{tab:sota}}
    \label{tab:after_gtod}
\end{table}

\section{Everest}\label{sec:Topk}
\name~allows users to set a per-query threshold, 
$\thr$, to ensure that the returned Top-K result 
has a minimum of $\thr$ probability to be the exact answer.
Given an uncertain relation $D$ 
obtained from a video 
(Section \ref{sec:phase_1} will discuss how to obtain that)
and a scoring function $S$ (\added{e.g., defining the score of a frame $f$ be the number of cars in $f$ identified by an oracle object detector}), 
\name\ returns a Top-K result $\hat{R}$ 
with confidence $\hat{p} = \Pr (\hat{R} = R) \ge \thr$,
where $R$ is the exact result and $\thr$ is the probability threshold specified by the user.
The probability $\hat{p}$ is defined over an uncertain relation $D$ (e.g., Table \ref{tab:sota}) using the 
\emph{possible world semantic} (PWS) \cite{possible_world}.
The possible world semantic is widely used in uncertain databases, 
where an uncertain relation is instantiated to multiple possible worlds, each of which is associated with a probability. 
Table \ref{tab:possible_world} shows two possible worlds (out of $3^3$) of Table \ref{tab:sota}. 
Given the probability of each possible world, the confidence $\phat$ of a Top-K answer $\hat{R}$ is the sum of probabilities of all possible worlds in which $\rhat$ is Top-K:
\begin{equation}
    \hat{p} = \sum_{W \in \mathcal{W}(D) \land \hat{R} = \text{Top-K}(W)} \Pr(W).
    \label{eq:define_confidence}
\end{equation}
Here, $\mathcal{W}(D)$ denotes the set of all possible worlds of 
an uncertain relation $D$.
In addition, 
the answer $\rhat$ has to satisfy the following:

\begin{definition}
{\bf The Certain-Result Condition}:  
The Top-K result \emph{$\rhat$ has to be chosen from $D^c$}, where frames in $D^c \subseteq D$ are all {\bf certain}, i.e., 
their frame scores are obtained from \replaced{a ground-truth object detector (GTOD)}{the given oracle} so that they have no uncertainty.
\end{definition}

The certain-result condition is important and unique in video analytics.
For instance, 
the Top-1 result of the uncertain relation $D$ in Table \ref{tab:sota}
is $\rhat= \{ f_3 \}$ and it has a confidence of $\phat=0.85$ based on Equation \ref{eq:define_confidence}.  
Assuming $\thr=0.8$, the confidence of $\rhat$ being the correct answer is above the user threshold.

\begin{figure*}[t]
    \centering
    \includegraphics[width=\textwidth]{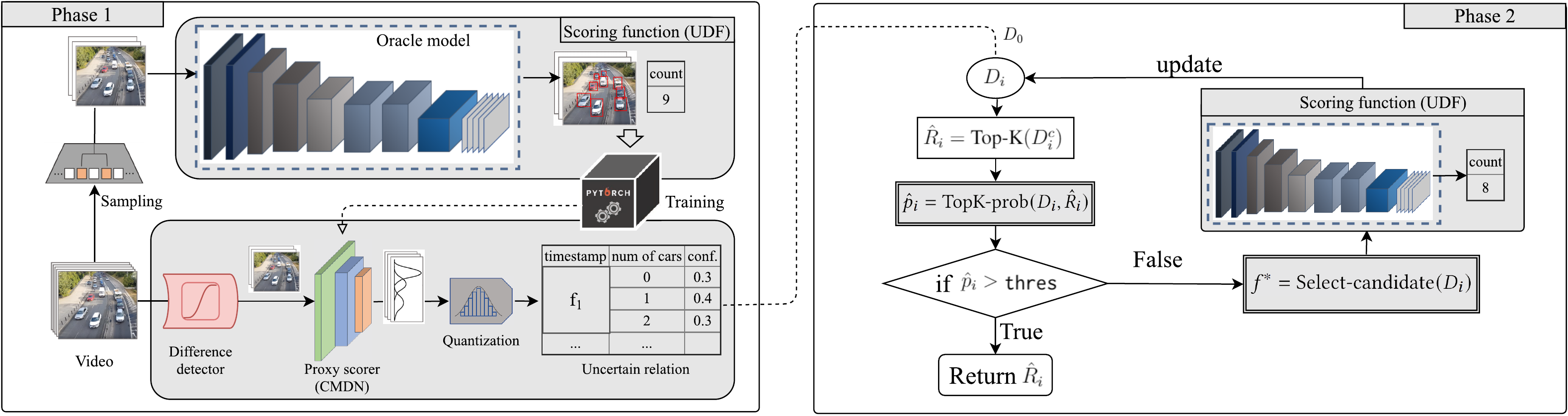}
    \caption{System Overview}
    \label{fig:overview}
\end{figure*}

However, after all this is a probabilistic result,
and so the users will feel weird if they 
visually inspect $f_3$ and actually see no cars in it.
The certain-result condition avoids such awkward answers and 
constrains that all the tuples in $\rhat$ have to be confirmed by \replaced{a GTOD}{the oracle} before it is returned to the users.
With the certain-result condition, $\{ f_3 \}$ would not be returned as the Top-1 result
because its probability of being Top-1 is only 0.38
based on the \textit{updated} uncertain relation $D'$ in Table \ref{tab:after_gtod}, in which the exact score of $f_3$ 
is obtained from \replaced{a GTOD}{the oracle} (we denote that operation as \replaced{GTOD($f_3$)}{\textsf{Oracle}($f_3$)}).
Finally, we remark that $\hat{p} = \Pr (\hat{R} = R) \ge \thr$
guarantees not only the whole result set $\hat{R}$ has at least $\thr$ probability of being the true answer, but every frame $\hat{f}$ in $\hat{R}$ also has at least $\thr$ probability of being in the exact result set $R$ because 
$\Pr(\hat{f} \in R) \ge \Pr(\hat{R} = R) \ge \thr$.
$\Pr(\hat{f} \in R)$ reflects the \textit{precision} of Top-K answer, i.e., the fraction of results in $\rhat$ that belongs to $R$. Therefore, \name~effectively provides probabilistic guarantees on the precision of the query answers.

\subsection{System Overview}\label{sec:system}
Following recent works \cite{tahoma,focus,noscope}, \name~also focuses on a batch setting.
In this setting, large quantities of video are collected for post-analysis. 
Online analytics on live video stream is a different setting and is beyond the scope of this paper.
\deleted{We focus on a fixed set of labels because 
the pre-trained GTOD 
can detect ``cars'' but 
cannot distinguish between sedan and hatchback.  
Users can supply {user-defined functions} (UDFs) to do further classification if necessary.}
To our knowledge, tracking model drift in visual data is still an ongoing research in computer vision \cite{noscope}. %
We will tackle that problem upon robust techniques for resolving model drift are developed.

Figure \ref {fig:overview} shows a system overview.
\name~leverages CNN specialization and uncertain query processing
to accelerate Top-K analytics with probabilistic guarantees.
Processing a query involves two phases.
The first phase trains a lightweight 
\emph{convolutional mixture density network} (CMDN) \cite{d2017DCMDN} \added{that outputs a rough score distribution for each frame to form an initial uncertain relation  $D_0$ quickly}.
The second phase takes as input the resulting uncertain relation $D_0$
from Phase 1
and finds a Top-K result $\rhat$ that has a confidence $\phat \ge \thr$.  
Initially, given $D_0$ only, 
it is unlikely that the initial Top-K result $\rhat_0$ from $D_0$ gives a confidence that is above the threshold.
Furthermore, given a potential Top-K result $\rhat$,
we have to confirm its frames for the certain-result condition using \replaced{a GTOD}{the oracle} --- but that 
may conversely give the same $\rhat$ a lower confidence based on the updated uncertain relation $D'$ (e.g., drops from 0.85 to 0.38).
Of course, if the uncertain relation $D'$ contains no more uncertainty (i.e., all tuples are certain),
the Top-K result from that $D'$ has a confidence of 1.
Consequently, Phase 2 
can be viewed as \textit{Top-K processing via online uncertain data cleaning},
in which the system selectively ``cleans'' the uncertain tuples in the uncertain relation using \replaced{a GTOD}{the oracle} until the Top-K result from the latest uncertain relation
satisfies the probabilistic guarantee.
For high efficiency, 
Phase 2 aims to clean as few uncertain tuples
as possible because each cleaning operation invokes the computationally expensive \replaced{GTOD}{oracle}.  Furthermore, the algorithms in Phase 2 have to be carefully designed because uncertain query processing often introduces an exponential number of possible worlds.

\subsection{Phase 1: Building the initial uncertain relation $D_0$
using convolutional mixture  density  network (CMDN)} \label{sec:phase_1}

The crux of CNN specialization is to 
design a fast but less accurate proxy model to approximate the functionality of the original accurate but slow-running \added{oracle model} (e.g., a less accurate but fast running object detector).
\added{Since users in \name~can specify different scoring (ranking) \RI{R1O1}
functions by providing any deep model as \added{the oracle} (e.g., sentimentalizer, depth estimator) at run-time, it is impossible
for \name~to prepare the proxy models ahead.
In view of this, we decide to build a 
\emph{proxy scorer} that can approximate any \emph{score distribution}
instead of building a proxy that approximates an oracle's  original functionality}.
In order to approximate any arbitrary scoring
function and predicting the probability density 
of a frame's estimated score (instead of a point-estimate),
we train a \emph{convolutional mixture density network} (CMDN) to approximate the score distribution.  
With that, the training data is 
obtained by 
randomly sampling frames from the video-of-interest
and obtaining their exact scores through invoking the accurate oracle.

\begin{figure}
    \centering
    \includegraphics[width=\linewidth]{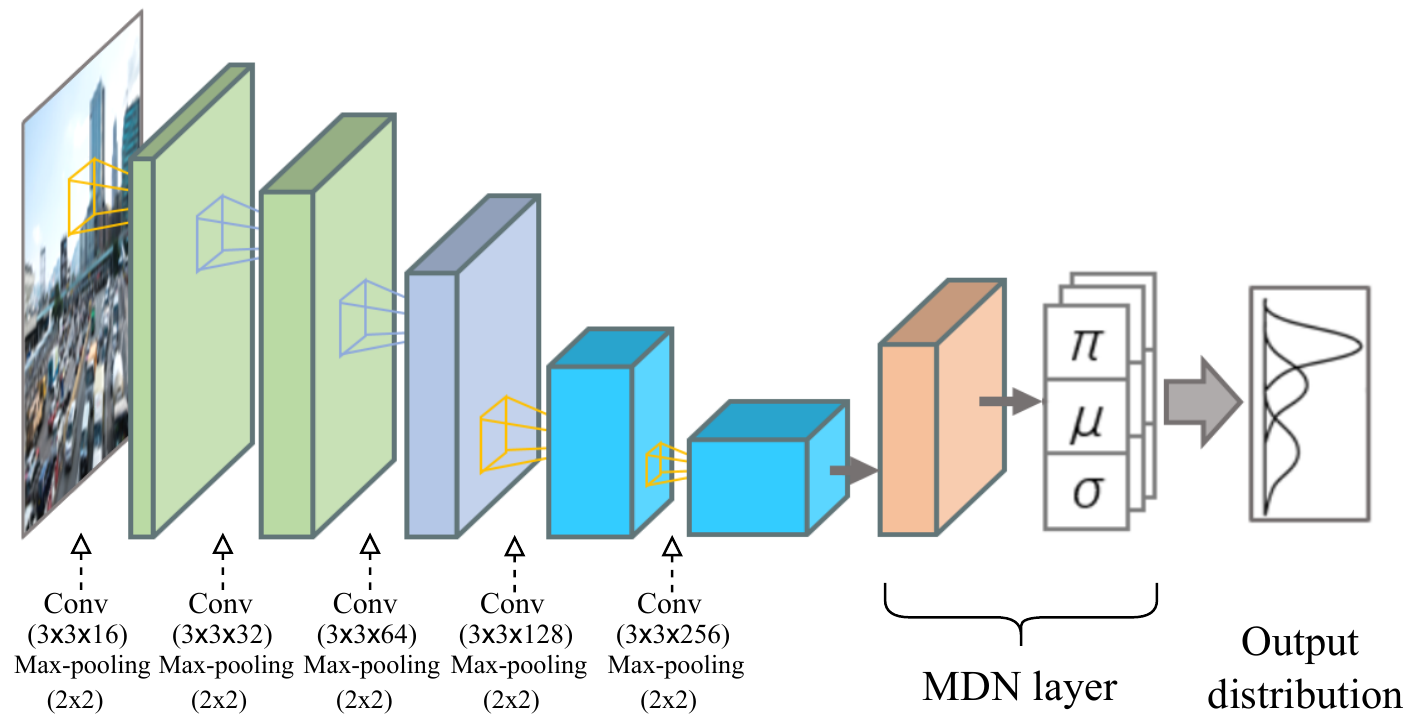}
    \caption{\added{\name~uses a Convolutional Mixture Density Network as a proxy scorer}}\label{fig:mdn}
    \RICAP{R1O3}
    \vspace{-0.5cm}
\end{figure}

Figure \ref{fig:mdn} shows the CMDN we designed. 
It uses five convolution layers to extract features from the input frame, \added{where the $i$-th layer has $2^{i+3}$ filters of $3\times 3$ kernel, \RI{R1O3}
followed by a $2\times2$ max-pooling}. 
Finally the extracted features are fed to a mixed density network (with $h$ hypothesis) to output parameters of $g$ Gaussians
(each has a mean $\mu$ and variance $\sigma$) and their weights $\pi$ in the mixture.

In order to generate robust models,
we train multiple CMDN models with different
sets of hyperparameters (e.g., $h$ and $g$).
Since both the model and the training data are small, 
the total training time is less than several minutes. 
After training, \name~selects the best model 
by evaluating the models on a holdout set.
The holdout set is obtained the same way as the training samples.
The model with the smallest \emph{negative log-likelihood} (NLL) \cite{nll} is chosen and the rest are discarded.

Before building the uncertain relation using the chosen CMDN, \name~uses a difference detector to discard similar frames from the video.
This step serves two purposes.
First, frames with little differences are not informative. %
Second, it approximates independence among frames so as to enable the use of ``x-tuples'' to model the data.
After that, \name~feeds 
the unique frames to the CMDN to obtain their score distributions.
An x-tuple captures a discrete distribution
but the Gaussian mixture is continuous and with infinitely long tails on both ends.
In order to get a finite uncertain relation,
we follow \cite{truncated_gaussian} 
to truncate the Gaussians so that the probabilities beyond $3\sigma$ are set to zero 
and evenly distributed to the rest.
After that, we 
populate the uncertain relation $D_0$
by quantizing the truncated mixed Gaussian distribution.
For counting based scoring function, 
the score distribution is quantized to 
a discrete distribution with 
non-negative integer support.
For others, 
users have to provide the quantization step size  when 
defining the scoring function.
For frames whose exact scores were obtained
during the collection of training/holdout data, 
they are inserted into the uncertain relation
straight with no uncertainty so that no work is wasted.
Nevertheless, we still call that as an ``uncertain relation''.

\subsection{Phase 2: Top-K processing with
ground-truth-in-the-loop uncertain data cleaning}\label{un}\label{sec:phase2}

\added{Given the
uncertain relation $D_0$ initialized in Phase 1, 
the second phase aims to locate a Top-K result $\rhat$ 
whose confidence $\phat$ exceeds $\thr$.
As mentioned, the initial $D_0$ populated by the CMDN 
is likely to contain too much uncertainty 
that no Top-K result obtained from it can pass the threshold.
Hence, starting with $D_0$,
the second phase 
(see Figure \ref{fig:overview} (right))
iteratively selects the best frame $f^*$ 
(by the {\sf Select-candidate} function)
to clean using the \replaced{GTOD}{oracle}.
This process is repeated
until a Top-K result $\rhat_i$ 
obtained from the latest table $D_i$
has a confidence $\phat_i$ 
(computed by the {\sf Topk-prob} function)
that exceeds $\thr$.
}

Finding a Top-K result $\rhat_i$ 
from the latest updated table $D_i$
(by the {\sf Top-K($D_i^c$)} function) is straightforward
because it simply extracts all the {\bf certain} tuples $D_i^c$ (because of the certain-result condition)
from $D_i$ and applies an existing Top-K algorithm (e.g., \cite{saying-enough}) to find the Top-K as $\rhat_i$.
The function {\sf Topk-prob} that computes the probability $\phat_i$ for
$\rhat_i$ 
being the exact result
and the function {\sf Select-candidate} that selects the most promising frame 
are more challenging because 
they involve an exponential number of possible worlds.
\added{Unfortunately, 
there are 
no existing techniques 
that can 
use an oracle to process uncertain Top-k queries with probabilistic guarantees.}

\deleted{Traditional uncertain data cleaning problems have a budget constraint,
which is usually the number of x-tuples that could be cleaned \cite{clean-topk, cheng2008cleaning, eric-ee}.  Their objective is to minimize the \emph{entropy} of all possible Top-K results from the updated table \cite{clean-topk, davidson2013using}.
Since their setting is to return a batch of x-tuples for manual offline cleaning,
using entropy allows them to measure the answer quality solely based on the uncertain data as 
\emph{the results of the cleaning operation
are not available at run-time}.
By contrast, our constraint is to pass the probability threshold $\thr$
and our objective is to minimize the cost of using the GTOD (and the algorithm overhead).
For us, the GTOD provides \emph{instant feedback},
allowing us to put that in the loop to come up with a probabilistic guarantee (between 0 and 1), which is way more intuitive than using entropy (which could be 0 to $\infty$).
Recent approaches that are based on crowdsourcing to 
clean uncertain data online 
while processing a Top-K query (e.g., \cite{davidson2013using,kou2017crowdsourced,li2018rating,ciceri2015crowdsourcing})
are also inapplicable 
because they are \emph{heuristic solutions}
and \emph{stop once the given monetary budget is exhausted}.
They are thus hard, if not impossible, to 
establish our probabilistic guarantee.}

\begin{table}\small
\centering
\begin{tabular}{|c||p{6.3cm}|}\hline
$D$ & Uncertain relation \\\hline
$D_i$ & Uncertain relation at iteration $i$ \\\hline
$D^c_i$ & Subset of $D_i$ whose x-tuples are all certain \\\hline
$D^u_i$ & Subset of $D_i$ whose x-tuples are all uncertain \\\hline
$\rhat$ & Approximate Top-K result \\\hline
$\phat$ & Confidence/probability of $\rhat$ \\\hline
$\rhat_i, \phat_i$ & $\rhat$, $\phat$ obtained in $i$-th iteration \\\hline
$R$  & Actual result \\\hline
$f$ & Frame / x-tuple \\\hline
$S_f$ & Score of $f$\\\hline
$\tf$ & The frame ranked $K$-th in $\rhat_i$ \\\hline
$\pf$ & The frame ranked penultimately in $\rhat_i$ \\\hline

\end{tabular}
\caption{Major Notations}
\label{tbl:notation}
\end{table}

In the following,
we discuss efficient algorithms to 
implement the {\sf Topk-prob} and 
the {\sf Select-candidate} functions.
Table \ref{tbl:notation} gives a summary of the major notations used.

\subsubsection{{\sf Topk-prob}($D_i$, $\rhat_i$)} \label{sec:no_gap_topk_prob}
Given a potential result $\rhat_i$ extracted from $D_i^c$,
computing its confidence $\phat_i$ via Equation \ref{eq:define_confidence} 
has to expand all $O(m^n)$ possible worlds, 
where $n$ is the number of frames in $D$ and assume each of them has $m$ possible scores in the uncertain relation.
However, given the certain-result condition, 
we can simplify the calculation of $\phat_i$ as:

\begin{equation}
    \phat_i =  \prod_{f \in D_i^u} \Pr(S_f \le S_{\tf}) 
    \label{eq:topk_prob}
\end{equation}

\noindent
where (a) $S$ stands for the score of a frame, 
(b) $\tf$ is the ``threshold'' frame that ranks $K$-th in $\rhat_i$, 
and its score $S_\tf$ is known and certain (because $\tf$ is from $\rhat_i \subseteq D_i^c$),
and (c) $D_i^u$ are frames in $D_i$ with uncertainty, i.e., $D_i^u = D_i \setminus D^c_i$.

Computing Equation \ref{eq:topk_prob} requires only time that is linear to $|D_i^u|$.
Equations \ref{eq:define_confidence} and \ref{eq:topk_prob} are equivalent
because the probability of $\rhat_i$ being Top-K is equal to the probability that 
no frames in $D_i^u$ having scores larger than the frames in $\rhat_i$.\footnote{We allow frames in $D_i^u$ to have scores tie with the threshold frame $\tf$.}

A further optimization is to compute two functions 
before Phase 2 begins:
(a) the CDF $F$ for the score of each frame $f$,
i.e., $F_f(t) = \Pr(S_f \le t) = \sum\limits_{j=0}^t \Pr(S_f = t)$ 
and (b) a function $H(t) = \prod_{f \in D_0^u} F_f(t)$,
which is the joint CDF of all uncertain frames in $D_0$.
With them, in the $i$-th iteration, 
we can compute $\phat_i$ as follows:

\begin{equation}
     \phat_i =\frac{H(S_{\tf})}{\prod_{f\in D_i^c} F_f(S_{\tf})}
     \label{eq:topk_prob_improve}
\end{equation}

Equations \ref{eq:topk_prob} and \ref{eq:topk_prob_improve} 
are equivalent 
because by definition $D_i^u = D_0^u \setminus D_i^c$.
$F_f(t)$  and $H(t)$ for all $f$ and $t$
can be easily pre-computed one-off at a cost of $O(|D_0^u|)$.  
With Equation \ref{eq:topk_prob_improve}, {\sf Topk-prob($D_i$, $\rhat_i$)} in the $i$-th iteration can compute $\phat_i$
using $O(|D_i^c|)$ time instead, where $|D_i^c| << |D_i|$.

According to Equation \ref{eq:topk_prob},
$\phat_i$ improves
exponentially with the number of frames cleaned.
Therefore, we expect Phase 2 would 
spend more iterations to reach a small probability threshold, say, 0.5. 
But after that, it would take fewer iterations to reach any probability threshold beyond.

\subsubsection{{\sf Select-candidate}($D^u_i$)} \label{sec:select_nogap}

{\sf Select-candidate}($D^u_i$) is the function 
to select a frame $f^*$ from the set of uncertain frames $D^u_i$ 
in the $i$-th iteration 
to obtain the exact score using the \replaced{GTOD}{oracle} 
such that cleaning $f^*$ can maximize  $\phat_{i+1}$
of the next iteration; 
and hopefully $\phat_{i+1} \ge \thr$ after that
and thus Phase 2 can stop early.

Of course, $\phat_{i+1}$ is unknown before we apply \replaced{GTOD}{\textsf{Oracle}}($f^*$).
Therefore, we use a random variable $X_f$ to denote the value of $\phat_{i+1}$ 
after a frame $f$ is cleaned. 
To maximize $\phat_{i+1}$, 
we aim to find $f^* = \arg\max_{f \in D_i^u}E_i[X_f]$. 
Using $x_f^s$ to denote the value of $X_f$ when $S_f=s$, 
where $s$ is a particular score, 
$E[X_f]$ is thus: 
\begin{equation}
\label{eq:expected_p}
  E[X_f] = \sum_{s} \Pr(S_f=s)x_f^s
\end{equation}

\stitle{[Efficient computation of $x_f^s$]}
$x_f^s$ is the probability of the result $\rhat_{i+1}$ 
being the Top-K based on $D_{i+1}$, 
where the x-tuple representing $f$ in $D_{i+1}$
is assumed to be cleaned and its score is $s$ and certain. 
Therefore, $x_f^s$ can be calculated on top of
$p_i$ (Equation \ref{eq:topk_prob_improve})
by removing the uncertainty of $f$, based on 
how the actual score $s$ of $f$ influences the Top-K result:

\begin{equation}
\label{eq:xfs}
  x_f^s = \left \{
  \begin{aligned}
     &\frac{H(S_{\tf})}{F_f(S_{\tf})\prod_{f'\in D_i^c} F_{f'}(S_{\tf})}=\frac{\phat_i}{F_f(S_{\tf})}  && s\le S_{\tf} \\
&\frac{H(s)}{F_f(s)\prod_{f'\in D_i^c} F_{f'}(s)} && S_{\tf} < s\le S_{\pf} \\
    &\frac{H(S_{\pf})}{F_f(S_{\pf})\prod_{f'\in D_i^c} F_{f'}(S_{\pf})}  && s > S_{\pf}
  \end{aligned} \right.
\end{equation}

The idea of Equation \ref{eq:xfs} is that:

\begin{itemize}
    \item when $s \le S_\tf$, 
    frame $f$ is not qualified to be in Top-K;
    the Top-K result would not change, 
 and the threshold score is still $S_\tf$;
So, discounting the uncertainty of $f$ suffices.

\item when $S_\tf < s\le S_{\pf}$, 
frame $f$ enters the Top-K and 
but its score is lower 
than the penultimate frame $\pf$ in the Top-K result, 
i.e., the one ranks ($K-1$)-st,
so frame $f$ gets the $K$-th rank;
the new ``threshold'' score is changed to $s$; 

\item when $s > S_{\pf}$, 
frame $f$ enters the Top-K
with a score greater than 
the original penultimate frame, 
the new threshold frame is $\pf$, 
the new threshold score is changed to  $S_{\pf}$. 
\end{itemize}

Putting Equations \ref{eq:xfs} and  \ref{eq:expected_p} together, we get:

\begin{equation}
\label{eq:expected_p_improved}
\begin{split}
E[X_f] = \phat_i &
+ \sum_{s=S_\tf+1}^{S_{\pf}} \frac{\Pr(S_f=s)H(s)}{F_f(s)\prod_{f'\in D_i^c} F_{f'}(s)} \\
&+ \frac{(1-F_f(S_{\pf}))H(S_{\pf})}{F_f(S_{\pf})\prod_{f'\in D_i^c} F_{f'}(S_{\pf})}
\end{split}
\end{equation}

Equation \ref{eq:expected_p_improved} greatly reduces the cost of computing $E[X_f]$ compared to Equation \ref{eq:expected_p} because 
the summation sums only over the range from $(S_\tf+1)$ to $S_{\pf}$.\\

\stitle{[Finding $f^*$ by early stopping]}
To find $f^*$ in an iteration $i$,
a baseline implementation of {\sf Select-candidate($D^u_i$)} has to compute $E_i[X_f]$ \emph{for every frame} $f$ in $D^u_i$.
It is inefficient because $D^u_i$ is large.

Fortunately, we can deduce an 
upper bound, $U(X_f)$, for each $E[X_f]$ 
and process frames in descending order of $U(X_f)$ to early stop the computation of $f^*$.
Specifically, starting from
Equation \ref{eq:expected_p_improved}, we have:

\begin{alignat}{2}
E[X_f] &= \phat_i &&+ \sum_{s=S_\tf+1}^{S_{\pf}} \frac{\Pr(S_f=s)H(s)}{F_f(s)\prod_{f'\in D_i^c} F_{f'}(s)} \nonumber\\
& &&+ \frac{(1-F_f(S_{\pf}))H(S_{\pf})}{F_f(S_{\pf})\prod_{f'\in D_i^c} F_{f'}(S_{\pf})} \nonumber\\
&{\color{red}\le} \phat_i &&+ \sum_{s=S_{\tf}+1}^{S_{\pf}} \frac{\Pr(S_f=s)H({\color{red}S_{\pf}})}{F_f({\color{red}S_{\pf}})\prod_{f'\in D_i^c} F_{f'}({\color{red}S_{\pf}})} \nonumber\\
& &&+ \frac{(1-F_f(S_{\pf}))H(S_{\pf})}{F_f(S_{\pf})\prod_{f'\in D_i^c} F_{f'}(S_{\pf})} \nonumber\\
&= \phat_i &&+ \frac{(1-F_f(S_\tf))H(S_{\pf})}{F_f(S_{\pf})\prod_{f'\in D_i^c} F_{f'}(S_{\pf})} \nonumber\\
&= \phat_{i} &&+ \gamma \psi_i(f) = U(X_f) \label{eq:upperbound}
\end{alignat}

where the last line factors the terms into:
$\gamma = \frac{H(S_{\pf})}{\prod_{f' \in D_i^c} F_{f'}(S_{\pf})}$ and
$\psi_i(f)=\frac{1-F_f(S_\tf)}{F_f(S_{\pf})}$.

In the $i$-th iteration, the order of frames' upper bound is only determined by the ``sort-factor'' $\psi_i(f)$ because 
the frames share the same $\phat_i$ and $\gamma$.
This suggests {\sf Select-candidate($D^u_i$)} to
examine the frames 
in the descending order of their $\psi_i(f)$. 
When it examines a frame $f^-$ whose $U(X_{f^-})$ is smaller than any examined frame $f_{seen}$'s $E[X_{f_{seen}}]$,
{\sf Select-candidate($D^u_i$)} would stop early and return $f^*$ from those that have been examined.
However, since 
$S_\tf$ (and thus $U(X_f)$) changes with $i$, we might have to re-compute and re-sort the frames per iteration. To avoid this overhead, we further re-write Equation \ref{eq:upperbound} to:

\begin{equation}
\label{eq:upperbound_j}
    E[X_f] \le \phat_i + \gamma \psi_{\color{red}j}(f)
\end{equation}

\noindent
where $j \le i$. 
The inequality still holds because $\psi_j(f) \le \psi_i(f)$ by observing
$S_{\tfj} \le S_{\tf}$ and $S_{\pfj} \le S_{\pf}$.

With Equation \ref{eq:upperbound_j}, we can simply set $j=0$ so that we only need to compute $\psi_j(f)$ and sort frames in the first iteration. 
However, to balance between tighter bounds and 
efficiency, in the first 100 iterations, we set $j=\floor*{\frac{i}{10}}$, 
i.e., we update $\psi_j(f)$ and sort frames every $10$ iterations.  
For iterations thereafter, we update $\psi_j(f)$ whenever $S_{\tf}$ or $S_{\pf}$ change. 
The idea is that 
$S_{\tf}$ and $S_{\pf}$ change more in early iterations 
but are relatively stable afterwards.

\subsection{Top-K Windows}
Videos are spatial-temporal in nature
and thus, users may want to 
split a video into \emph{time windows} of finite size,
compute their scores,
and examine the Top-K ones.
For example, 
an urban planner may 
be interested in the Top-50 5-second 
windows, where the score of a window
is the average number of cars observed in its frames.

\name\ supports Top-K over 
\emph{tumbling windows} like the example given above.
Specifically, a video-of-interest is divided into consecutive non-overlapping time windows $w_1$, $w_2$,...,$w_n$, each of which contains $L$ frames.
The score of a window $w$, denoted by $S_{w}$, is 
the average of the scores of the frames in it, 
i.e., $S_{w} = \frac{1}{L}\sum_{f \in w} S_f$.
For Top-K-window queries, we find the Top-K windows $\rhat$ 
such that $\phat = \Pr(\rhat=R) \ge {\tt thres}$, 
where $R$ is the set of true Top-K windows.

To support this type of query, 
\name~builds another uncertain relation whose schema is akin to Table \ref{tab:sota}: 
{\tt (window, avg(num of cars), prob)}.
Let the $i$-th frame in $w$ be $f^i$. 
The distribution of $S_{w}$ can be calculated based on the distributions of $S_{f^1}, S_{f^2}, \ldots, S_{f^L}$.
From Section \ref{sec:phase_1}, 
the distribution of $S_{f^i}$, as obtained from the CMDN, is a $g$-component Gaussian mixture 
with a density of: 

$$\sum_{j=1}^g \pi_{ij}\mathcal{N}(\mu_{ij}, \sigma_{ij}^2)$$

\noindent
where $\pi_{ij}$, $\mu_{ij}$, $\sigma_{ij}$ are the weight, mean and variance of the $j$-th component in the mixture distribution of $S_{f^i}$, respectively. 
Since \name's difference detector (Section \ref{sec:implementation}) discards 
a frame $f^i$ if it is too similar to a retained frame $r_t$, 
the score distribution of $S_{f^i}$ is approximated by the distribution of $S_{r_t}$.
Furthermore, \name~'s difference detector effectively divides a window into $l$ segments, where the frames in the same segment are similar to the same retained frame $r_t$.
Since the retained frames are judged by
the difference detector 
as sufficiently dissimilar, 
we assume their score distributions are independent.
Let $r_1$, $r_2$, ..., $r_l$ denote the $l$ retained frames in $w$,
we approximate the distribution of $S_w$ by

\begin{equation}
\label{eq:avg_approximate}
\begin{split}
        S_w & \sim \mathcal{N}(\frac{1}{L}\sum_{t=1}^l |s_t| \bar{\mu}_{r_t}, \frac{1}{L}\sum_{t=1}^l |s_t| \bar{\sigma}_{r_t}^2)
\end{split}
\end{equation}
where $|s_t|$ denotes the size of $t$-th segment; $\bar{\mu}_{r_t}$ and $\bar{\sigma}_{r_t}$ are the mean and the total variance of $S_{r_t}$, respectively. Suppose $r_t$ is the $q$-th frame in $w$, then

\begin{equation}
\begin{split}
    \bar{\mu}_{r_t} &= \sum_{j=1}^g \pi_{qj}\mu_{qj}, \\\nonumber
    \bar{\sigma}_{r_t}^2 &= \sum_{j=1}^g \pi_{qj} (\sigma_{qj}^2 + \mu_{qj}^2 - \bar{\mu}_{r_t}^2). \\\nonumber  
\end{split}
\end{equation}

By quantizing the distribution in Equation \ref{eq:avg_approximate}, \name~obtains an uncertain relation of x-tuples on the mean score of each tumbling window. That uncertain table is compatible 
with the algorithms in Phase 2.
When confirming a window using the \replaced{GTOD}{oracle}
(to compute the average score of a window)
during Phase 2, a large window size may require cleaning a lot of frames.
Therefore, we only sample some frames to verify with the \replaced{GTOD}{oracle} 
and compute the sample mean.

\subsection{System Details and Optimizations}\label{sec:implementation}
\name~is implemented in Python 3.7 with Numpy 1.17.
\deleted{As example oracle scorer, we use a Pytorch implementation of YOLOv3 \cite{yolo3} as the ground-truth object detector, whose weights are pretrained using the COCO dataset \cite{lin2014microsoft} with 416$\times$416 image resolution.}
Training and inference of the CMDN are implemented using PyTorch 1.4; and we use Decord 0.4 for video decoding.
\added{Currently \name~is a standalone system. 
\RII{R2O3}
One future work is to follow RAM${}^3$S \cite{forced_to_cite} 
to implement our techniques as a software framework 
so that we can leverage the various 
big data platforms (e.g., Spark) to scale-out.}

\lstset{basicstyle=\footnotesize\ttfamily,breaklines=true}
\lstset{frame=single}
\lstset{style=mystyle}
\begin{figure}
    \centering
\begin{lstlisting}[language=Python]
import ...
config = get_config_from_UI()
object_of_interest = config.obj
oracle = load_model("models/yolov3.pth")
def score_func(frames):
  object_lists = oracle(frames, object_of_interest)
  scores = [len(objects) for objects in object_lists] 
  return scores
\end{lstlisting}
    \caption{\added{An object counting UDF}}
    \vspace{-0.3cm}
    \begin{varwidth}[t]{1cm}
    \METACAP{MR2}
    \end{varwidth}
    \begin{varwidth}[t]{1cm}
    \RICAP{R1O1}
    \end{varwidth}
    \begin{varwidth}[t]{1cm}
    \RIIICAP{R4O1}
    \end{varwidth}
    \label{fig:udf}
    \vspace{-0.6cm}
\end{figure}

\stitle{CMDN Training.} 
The sampled frames are resized to 128$\times$128 resolution and pixel values are normalized to the range from 0 to 1. 
\name~trains $4\times3=12$ models of different hyperparameters and selects the best one with the smallest negative log-likelihood. The set of hyperparameters are $g=\{5, 8, 12, 15\}$
and $h=\{20, 30, 40\}$,
where $g$ is the number of Gaussians and 
$h$ is the number of hypotheses in the MDN layer. 
All models use five convolution layers 
because that is empirically stable across all videos 
and further reducing the number of convolution layers offers no additional
speedup because decoding the video would become the bottleneck.

\stitle{Difference Detector.} 
Various difference detectors (e.g., \cite{sysml18if}) can be used in \name.
In the current implementation, we follow \cite{noscope} and use mean-square-error (MSE) among pixels to measure the difference between two frames. 
To eliminate similar consecutive frames, in principle we need to 
sequentially scan through the video and discard a frame if its MSE with the last retained frame is lower than a threshold.
In order to parallelize this step, 
we split the video into clips of $c$ frames each. 
Each frame in a clip is compared with the middle one in the clip (i.e., the $\floor{\frac{c}{2}}$-th)
and is discarded if their MSE is lower than a threshold.
The clips are then processed in parallel. Although similar frames at the boundaries of clips may be retained or two adjacent clips may be quite similar, $c$ can be adjusted to ensure these situations are rare.

\stitle{Batch Inference.} 
The {\sf Select-candidate} function in Phase 2 
selects the most promising frame and infers its exact score using the \replaced{ground-truth object detector}{oracle} in each iteration. 
However, selecting and confirming
only one frame at a time might not fully utilize the abundant GPU processing power.
Therefore,
our implementation selects a batch of $b$ %
frames that have the highest expectations based on Equation \ref{eq:expected_p_improved}
and carries out batch inference.
The value of $b$ depends on the FLOPS and the memory bandwidth of GPU. Although the GPU is better utilized for larger $b$, setting $b$ too large may cause cleaning unnecessary frames. 
Therefore, we choose $b$ based on a measurement of inference latency 
to ensure that the latency of cleaning $b$ frames has no significant difference from cleaning one frame. 

\stitle{Prefetching.}Deep network inference may have I/O overheads 
when frames are fetched from the disk to the main memory 
thus stalling the GPU.
The baseline scan-and-test approach can alleviate that easily 
by prefetching the frames because it accesses 
them sequentially.  
Our Top-K algorithm, however, selects frames to clean, which is non-sequential.
Fortunately, \name~can 
achieve high throughput by also prefetching the input frames
based on the sort-order of $\psi_j$ in Equation \ref{eq:upperbound_j}.
Therefore, batches of frames with the highest $\psi_j$
would be pre-fetched while the GPU is carrying out computation.

\added{
\stitle{User-defined Function (UDF).} \META{MR2} \RI{R1O1} \RIII{R4O1}
\name~accepts user-defined scoring function in the form of a Python module. 
Each UDF follows a signature 
of taking images as input 
and returning their oracle scores as output.
Figure \ref{fig:udf} shows 
the default scoring function in \name. 
It uses 
a Pytorch implementation of YOLOv3 \cite{yolo3} as the oracle 
and the number of appearances of the query object as the score.
The weights of that model 
are pretrained using the COCO dataset \cite{lin2014microsoft} with 416$\times$416 image resolution.

}

\section{Evaluation} \label{sec:exp}
We performed experiments on an
Intel i9-7900X server with 64GB RAM and one NVIDIA GTX1080Ti GPU.
The server runs CentOS 7.0.

\stitle{Queries and Datasets.} 
Except for the last experiment that evaluates \name's capability of different scoring functions (Section \ref{sec:7}),
we evaluate \name~based 
on the Top-K object counting.
The first five rows of Table \ref{tbldata} 
shows the details of the real videos used in those experiments.
Among them, three of them were also used in prior work and we add two moving camera videos ({\tt Daxi-old-street}\footnote{\url{https://www.youtube.com/watch?v=z_mlibCfgFI}} and {\tt Irish-Center}\footnote{\url{https://www.youtube.com/watch?v=MXqKk4WEhsE}
}) collected from Youtube. 
The main object-of-interest varies. For example, {\tt Archie} 
and {\tt Taipei-bus} 
are traffic footages and so their main objects are cars. By default, K=50 and \thr=0.9.

We also include 
synthetic datasets generated by the latest Visual Road benchmark \cite{visualroad}. The synthetic datasets are used in Section \ref{sec:6}
to evaluate the impact of the number of objects that appear in the video since we cannot control the number of objects in real videos.

\begin{table}\small

\centering
\resizebox{\linewidth}{!}{
\begin{tabular}{|c||c|c|c|c     |c|}
\hline
\begin{tabular}[c]{@{}c@{}}{\sf Video}\\ {\sf (Used in / From)}\end{tabular} & \begin{tabular}[c]{@{}c@{}}{\sf Object-of-} \\ {\sf  interest}\end{tabular} & {\sf Resolution} & {\sf FPS} & \begin{tabular}[c]{@{}c@{}}{\sf \# of} \\ {\sf  frames}\end{tabular} & \begin{tabular}[c]{@{}c@{}}{\sf Length}\\ {\sf (hrs)}\end{tabular} \\ \hline\hline
\multicolumn{6}{|c|}{Object Counting (UDF) Default}\\\hline
{\tt Archie} (\cite{blazeit}) & car   & 1920$\times$1080 & 30 & 2130k & 19.7\\  \hline

{\tt Daxi-old-street}$^2$ & person & 1920$\times$1080 & 30 & 8640k & 80\\ \hline

{\tt Grand-Canal} (\cite{noscope,blazeit}) & boat & 1920$\times$1080 & 60 & 25100k & 116.2\\ \hline

{\tt Irish-Center}$^3$ & car & 1920$\times$1080 & 30 & 32401k & 300\\ \hline

{\tt Taipei-bus} (\cite{noscope,blazeit}) & car & 1920$\times$1080 & 30 & 32488k & 300.8 \\ \hline\hline
\multicolumn{6}{|c|}{\added{Tailgating Degree (UDF) on Dashcam Videos}}\\\hline
{\tt Dashcam-California}\tablefootnote{\url{https://www.youtube.com/watch?v=eoXguTDnnHM}}  & car & 1280$\times$720 & 30 & 324k & 3 \\ \hline

{\tt Dashcam-Greenport}\tablefootnote{\url{https://www.youtube.com/watch?v=-fuNmR5e19o}} & car & 1280$\times$720 & 30 & 350k & 3.2 \\ \hline

\end{tabular}}
\caption{Dataset Characteristics}
\label{tbldata}

\end{table}

\added{
\stitle{Baselines.}
To our best knowledge, 
\name~is the first system that supports Top-K queries in modern video analytics. Therefore, there are no other systems
that we can directly compare \name~with.
Alternatively, we considered the following baselines in addition to the naive scan-and-test approach:

\begin{itemize}
    \item HOG \cite{hog} \META{MR1}\RII{R2O4}\RIII{R4O5} is a classic computer vision method that is able to count objects in video frames 
{\bf without using deep learning}.
HOG scans over hundreds of sub-regions of the input image and performs classifications on each of them using SVM with low-level features (e.g., gradients of pixels).
In this baseline, we scan the video using HOG and report the Top-K frames with the highest number of object-of-interest.

\item CMDN-only.
\RI{R1O2}
In this baseline, 
we consider only Phase 1 of \name~and rank a frame
using its \emph{mean} score from its distribution produced by the CMDN.  

\item TinyYOLOv3-only.
\RI{R1O2}
TinyYOLOv3 \cite{yolo3} is the light version of YOLOv3 for \emph{real-time} object detection. 
In this baseline, we scan the video using TinyYOLOv3 and look for the Top-K frames there.

\item Select-and-Topk.
Lastly, we consider a baseline that is based 
on recent systems that support selection queries (e.g., \cite{noscope, focus, tahoma}).
Specifically, we rewrite a Top-K query 
as a range selection query followed by a Top-K operation.
First, we issue a range query 
``$S_f \ge \lambda M$'' to such a selection system
to retrieve all frames $C$ with scores higher than 
$\lambda M$, where $M$ is the maximum score found 
during specialized CNN training
and $\lambda \in [0,1]$.
Then, $C$ is regarded as the set 
of candidate frames with the highest scores 
and the Top-$K$ in $C$ is returned as the answer.
We refer to this baseline as the \emph{Select-and-Topk} method.
This baseline, however, is impractical 
because it is hard to get the value 
$\lambda$ right.
For example, setting $\lambda$ too large may make the number of candidate frames $|C|$ smaller than $K$.
On the other hand, setting $\lambda$ too small would severely increase the selection latency because it gets a bigger set of candidate frames $C$. 
In our implementation of ``Select-and-Topk'', we choose NoScope \cite{noscope} to handle the range selection operation
because it is open-source.
We use the default parameters of NoScope's optimizer and set its tolerable false negative rate to 0.1
(to mimic \name's \thr=0.9)  
and false positive rate to 0 (to mimic \name's certain-result condition).
As there are two newer systems (Focus \cite{focus} and TAHOMA \cite{tahoma})
that are similar to NoScope, 
we give advantages to this baseline.
First, we manually calibrate $\lambda$ in each experiment 
and report the one that yields the largest speedup subject to precision over 0.9.
Second, we exclude NoScope's specialized CNN training time
because Focus \cite{focus} advocates to 
do training/indexing offline during data ingestion rather than online query processing.
Third, we count only its time spent on \replaced{the GTOD}{the oracle} in order to rule out the 
other factors such as the speed difference  
of using different video decoders.

\end{itemize}

}

\stitle{Evaluation Metrics.}
For each experiment, 
we report (a) the end-to-end query runtime (for \name, we include everything from Phase 1 to Phase 2 and the algorithm runtime) and the speedup over the naive scan-and-test method.
We also report the
result quality in terms of
(b) \emph{precision} (the fraction of results in $\rhat$ that belongs to $R$)\footnote{
The \textit{recall} (the fraction of results in $R$ that were covered by $\rhat$)
is the same as the precision because both $R$ and $\rhat$ contain $K$ elements.},
(c) \emph{rank distance} (the normalized footrule distance between the ranks of the $\rhat$ and their true ranks in $R$),
and 
(d) \emph{score error} (the average absolute error for scores between $\rhat$ and $R$).

\stitle{System configurations.} 
For \name~, in Phase 1, the number of frames in training data is set to $\min\{0.5\%n, 30000\}$, where $n$ is the number of frames. We cap the sample size to be 30000 
because that is sufficient even for the longest video (300.8 hours) in our datasets.
The size of the holdout set is 3000 frames for all the datasets.
Although the MSE threshold in the difference detector 
can be tuned based on the speed of the moving objects
or the speed of the moving camera,
we are able to 
use a unified MSE threshold of 0.0001 and clip size of 30 for all datasets.
In Phase 2, we set the batch inference size $b$ to be 8 after measuring the inference latency on our server.

\begin{table*}[]
\centering
\subfloat[Latency breakdown]{
\begin{tabular}[b]{c|c|c|c||c|c|}
\cline{2-6}
& \multicolumn{3}{c||}{{\sf Phase 1}} & \multicolumn{2}{c|}{{\sf Phase 2}}  \\ \hline
\multicolumn{1}{|c||}{{\sf Dataset}}    &  \multicolumn{1}{c|}{\begin{tabular}[c]{@{}c@{}}{\sf Label sample} \\ {\sf  frames by oracle}\end{tabular}} & \multicolumn{1}{c|}{\begin{tabular}[c]{@{}c@{}}{\sf CMDN} \\ {\sf training}\end{tabular}} & \multicolumn{1}{c||}{\begin{tabular}[c]{@{}c@{}}{\sf Populate $D_0$ by } \\ {\sf CMDN inference}\end{tabular}} &  \multicolumn{1}{c|}{\begin{tabular}[c]{@{}c@{}}{\sf Select-} \\ {\sf candidate}\end{tabular}} & \multicolumn{1}{c|}{\begin{tabular}[c]{@{}c@{}}{\sf Confirm frames} \\ {\sf using orale}\end{tabular}}  \\ \hline \hline
\multicolumn{1}{|l||}{{\tt Archie}} & 10.48\% & 35.19\% & 41.75\% & 0.16\% & 12.42\%  \\ \hline
\multicolumn{1}{|l||}{{\tt Daxi-old-street}} & 6.34\% & 43.48\% & 45.19\% & 0.11\% & 4.88\% \\ \hline
\multicolumn{1}{|l||}{{\tt Grand-Canal}} & 2.27\% & 20.82\% & 65.79\% & 0.34\% & 10.79\%  \\ \hline
\multicolumn{1}{|l||}{{\tt Irish-Center}} & 1.81\% & 15.81\% & 79.38\% & 0.16\% & 2.84\%  \\ \hline
\multicolumn{1}{|l||}{{\tt Taipei-bus}} & 1.87\% & 17.25\% & 72.23\% & 0.41\% & 8.24\% \\ \hline
\end{tabular}}
\subfloat[More about Phase 2]{
\begin{tabular}[b]{|c|c|}
\multicolumn{1}{c}{} & \multicolumn{1}{c}{}\\
\hline
\begin{tabular}[c]{@{}c@{}}{\sf Num of} \\ {\sf iterations}\end{tabular} & \begin{tabular}[c]{@{}c@{}}{\sf \% of frames} \\ {\sf cleaned}\end{tabular} \\ \hline \hline
2021 & 0.76\% \\ \hline
3173 & 0.29\% \\ \hline
19612 & 0.63\% \\ \hline
6474 & 0.16\% \\ \hline
18184 & 0.45\% \\ \hline 
\end{tabular}
}
\caption{A Detailed Breakdown}
\label{tab:breakdown}
\end{table*}

\begin{figure*}
    \begin{subfigure}{0.48\linewidth}
    \centering
    \includegraphics[width=\linewidth]{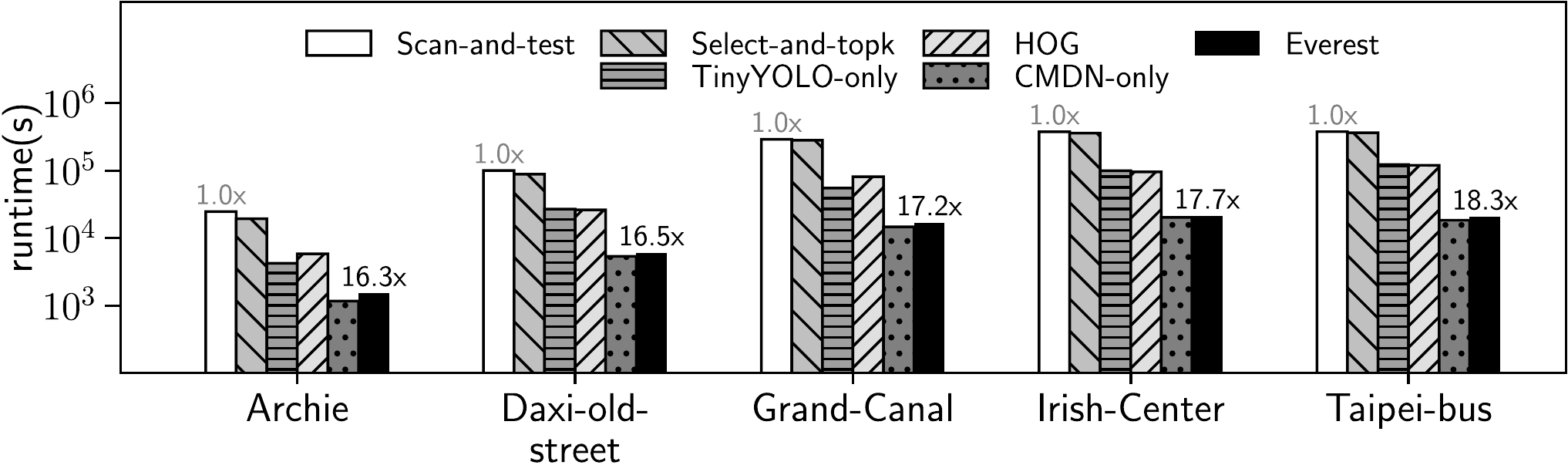}
    \caption{Speedup (log-scale)}
    \label{fig:my_label}
    \end{subfigure}
    \hfill
    \begin{subfigure}{0.48\linewidth}
    \centering
    \includegraphics[width=\linewidth]{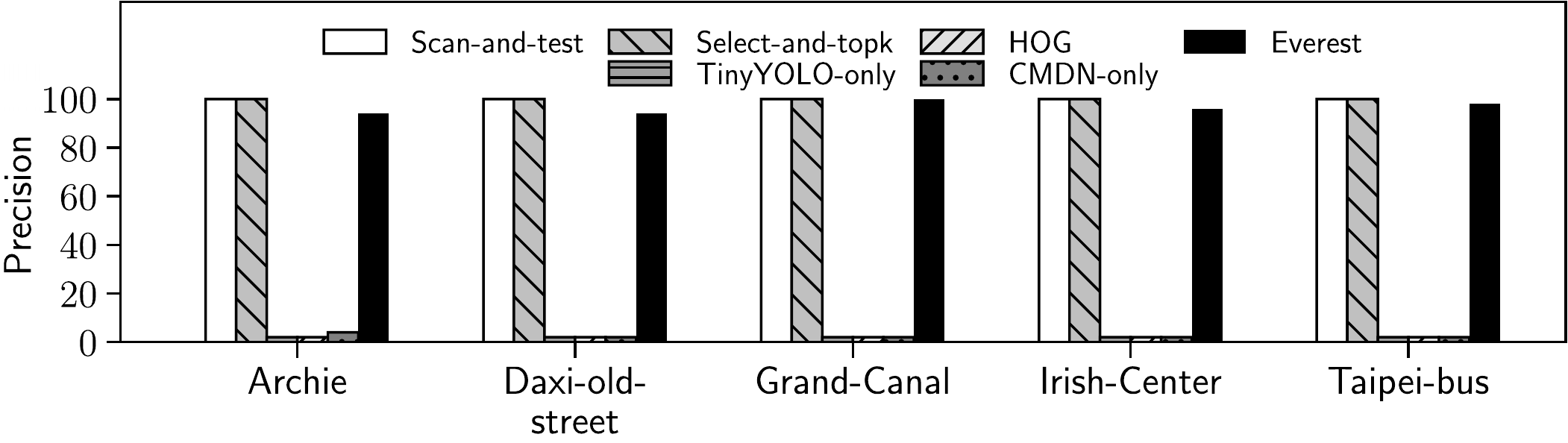}
    \caption{Precision}
    \label{fig:my_label}
    \end{subfigure}

    \vspace{2mm}
    \begin{subfigure}{0.48\linewidth}
    \centering
    \includegraphics[width=\linewidth]{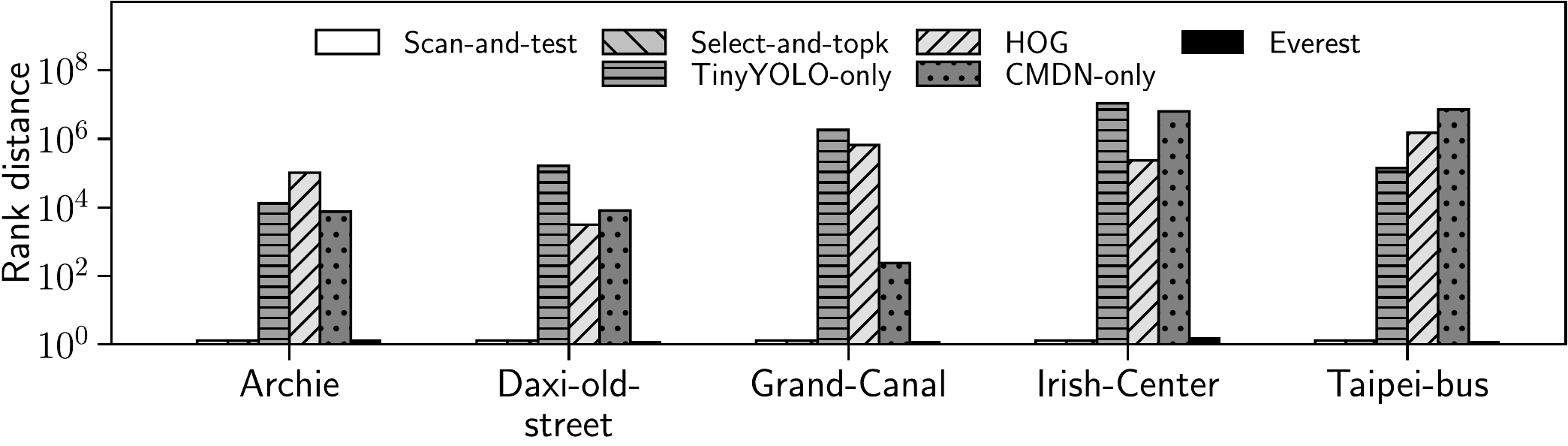}
    \caption{Rank distance (log-scale)}
    \label{fig:my_label}
    \end{subfigure}
    \hfill
    \begin{subfigure}{0.48\linewidth}
    \centering
    \includegraphics[width=\linewidth]{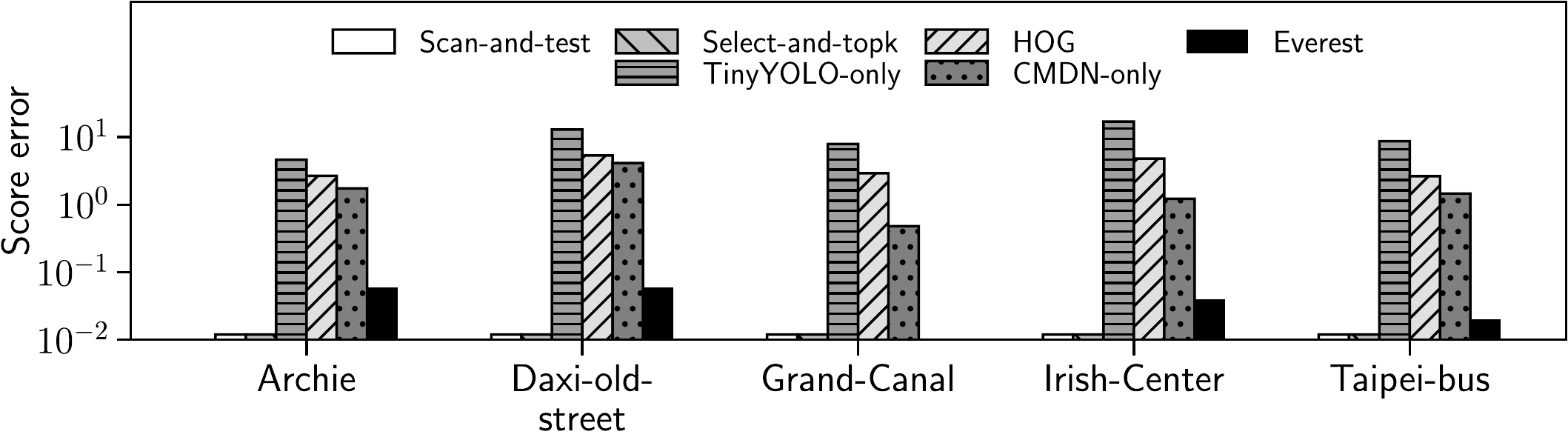}
    \caption{Score error (log-scale)}
    \end{subfigure}
    \caption{\added{Overall result under the default setting}}
    \begin{varwidth}[t]{1cm}
    \METACAP{MR1}
    \end{varwidth}
    \begin{varwidth}[t]{1cm}
    \RICAP{R1O2}
    \end{varwidth}
    \begin{varwidth}[t]{1cm}
    \RIICAP{R2O4}
    \end{varwidth}
    \begin{varwidth}[t]{1cm}
    \RIIICAP{R4O5}
    \end{varwidth}
    \label{fig:1}
\end{figure*}

\begin{figure*}
    \begin{subfigure}{0.245\linewidth}
    \centering
    \includegraphics[width=\linewidth]{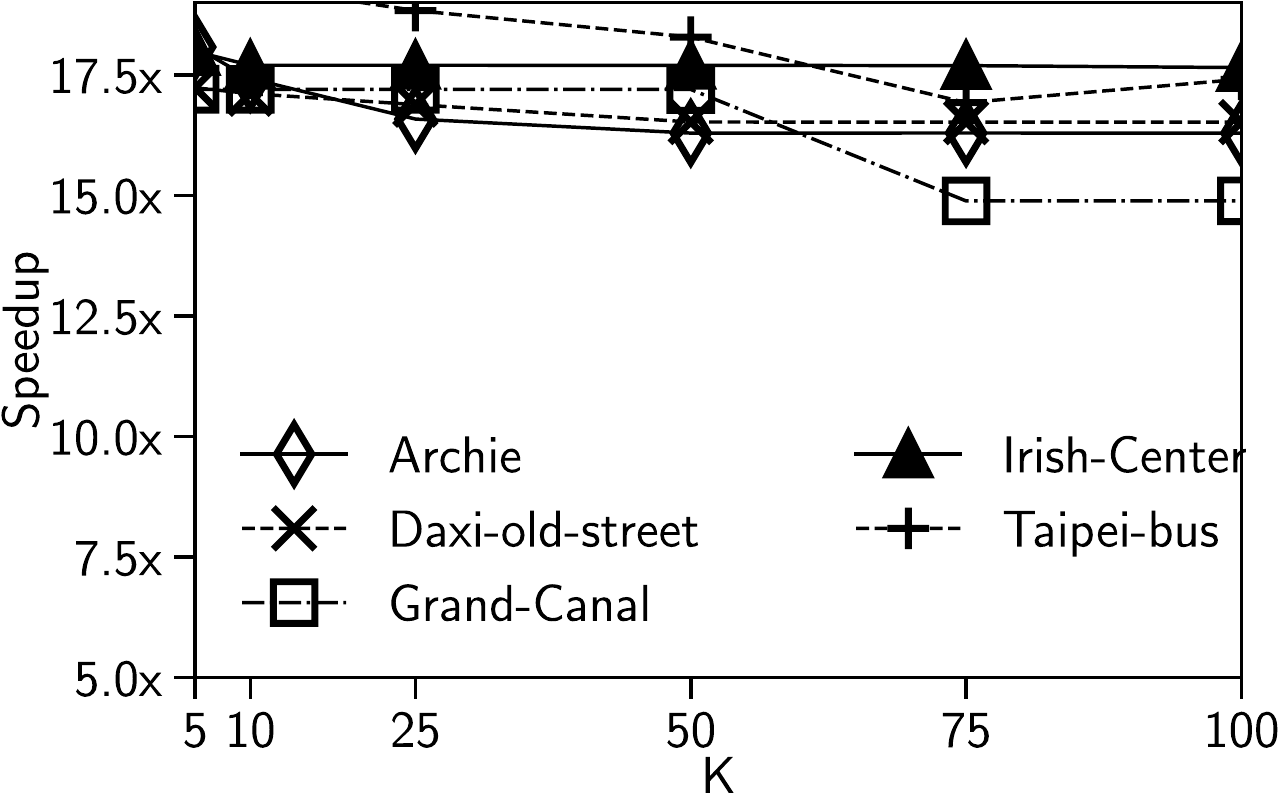}
    \caption{Speedup}
    \label{fig:my_label}
    \end{subfigure}
    \hfill  
    \begin{subfigure}{0.245\linewidth}
    \centering
    \includegraphics[width=\linewidth]{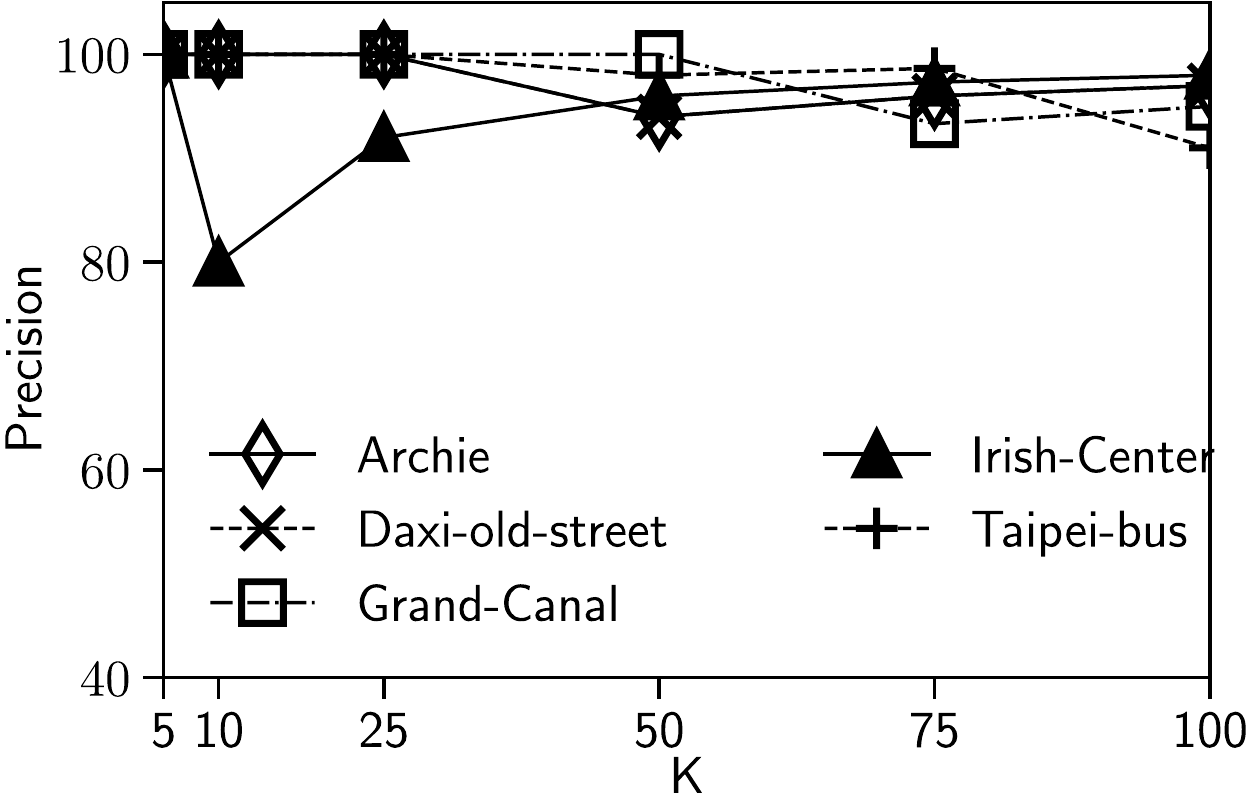}
    \caption{Precision}
    \label{fig:my_label}
    \end{subfigure}
    \hfill
    \begin{subfigure}{0.245\linewidth}
    \centering
    \includegraphics[width=\linewidth]{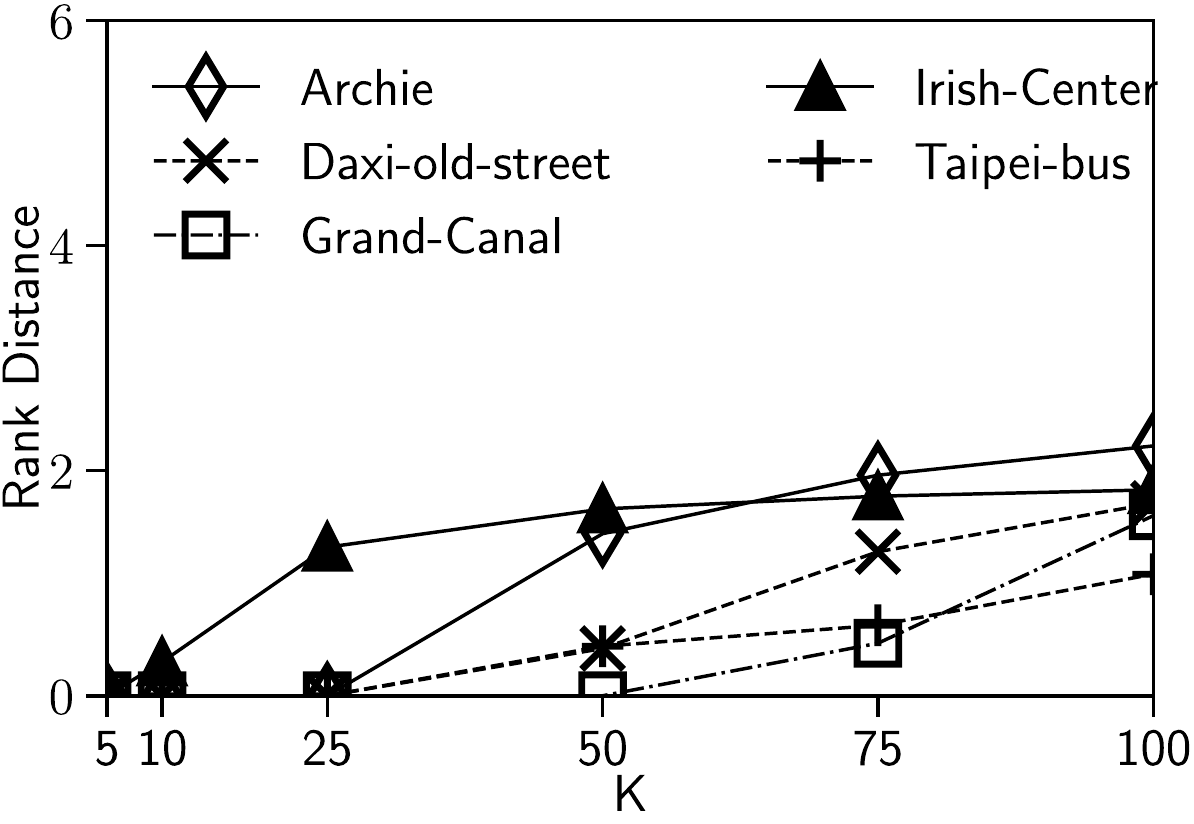}
    \caption{Rank distance}
    \label{fig:my_label}
    \end{subfigure}
    \hfill
    \begin{subfigure}{0.245\linewidth}
    \centering
    \includegraphics[width=\linewidth]{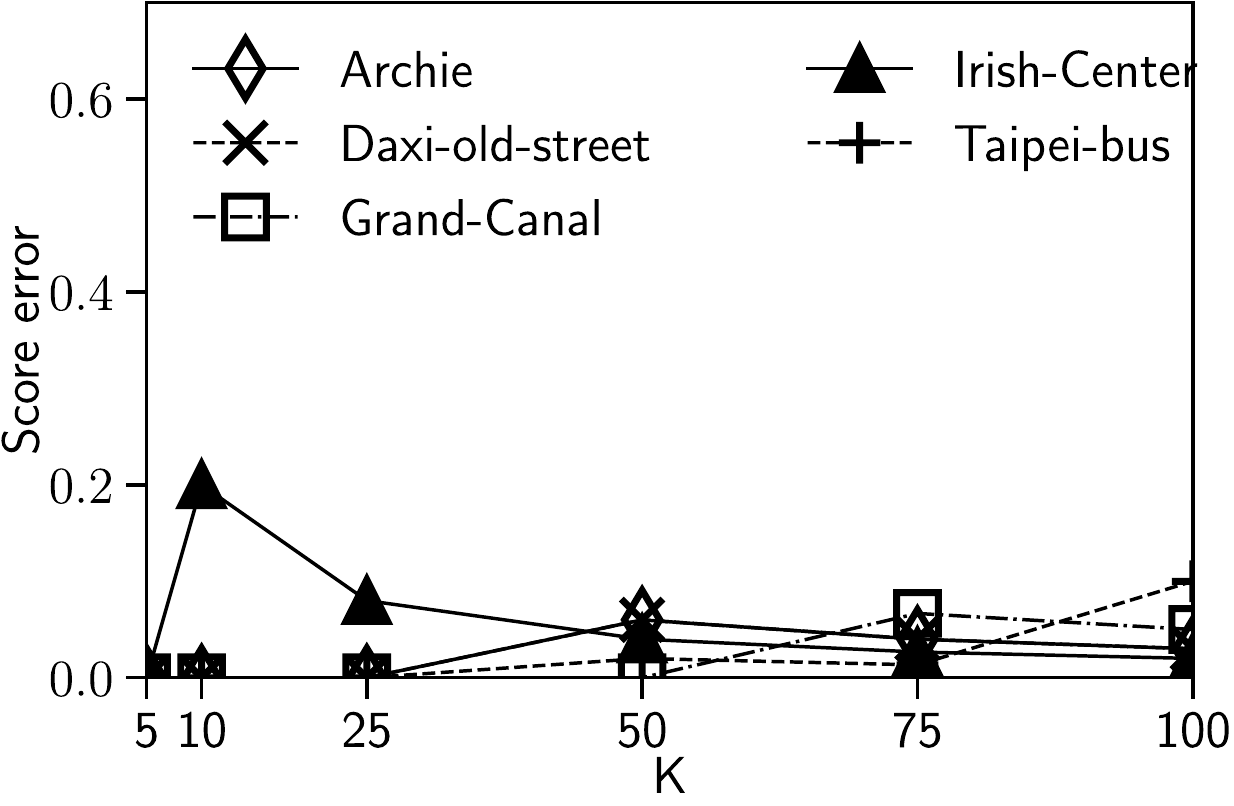}
    \caption{Score error (log-scale)}
    \end{subfigure}
    \caption{Impact of K}
    \label{fig:2}
\end{figure*}
\begin{figure*}
    \begin{subfigure}{0.245\linewidth}
    \includegraphics[width=\linewidth]{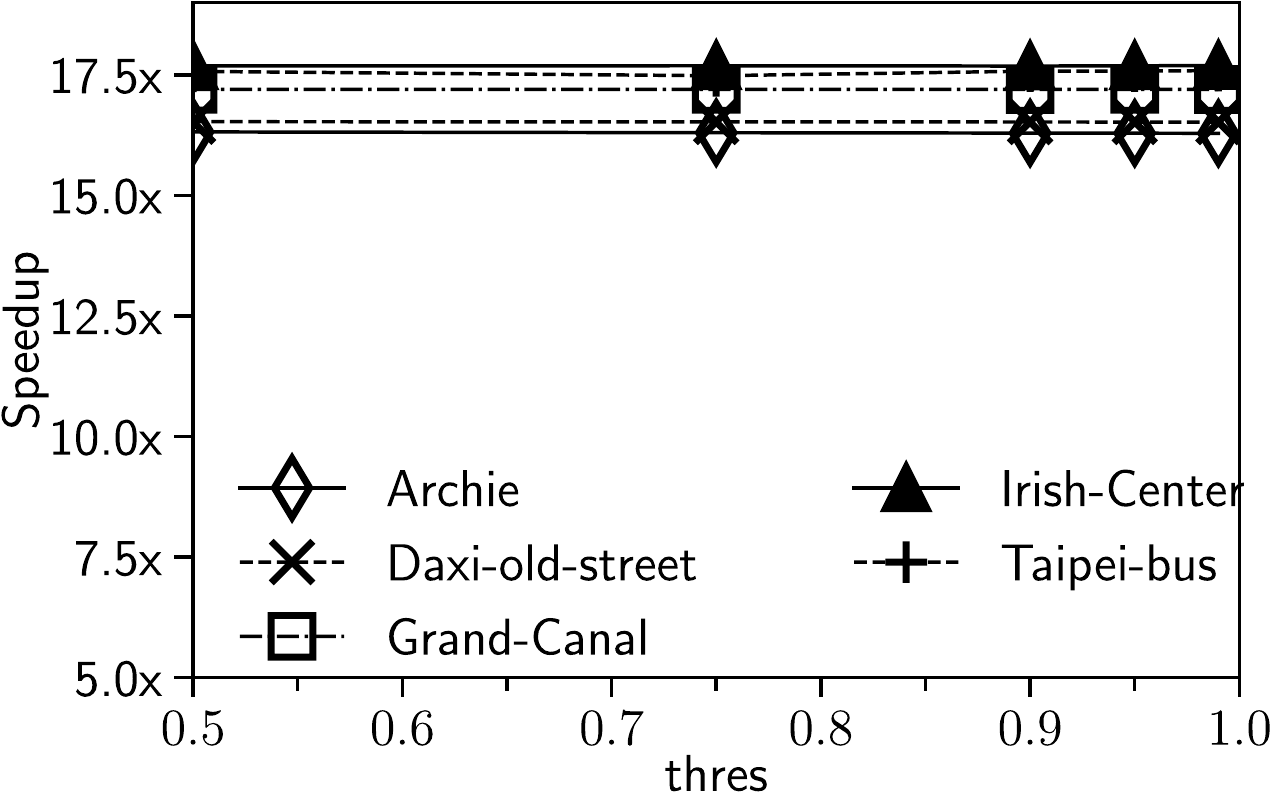}
    \caption{Speedup}
    \label{fig:my_label}
    \end{subfigure}
    \hfill
    \begin{subfigure}{0.245\linewidth}
\includegraphics[width=\linewidth]{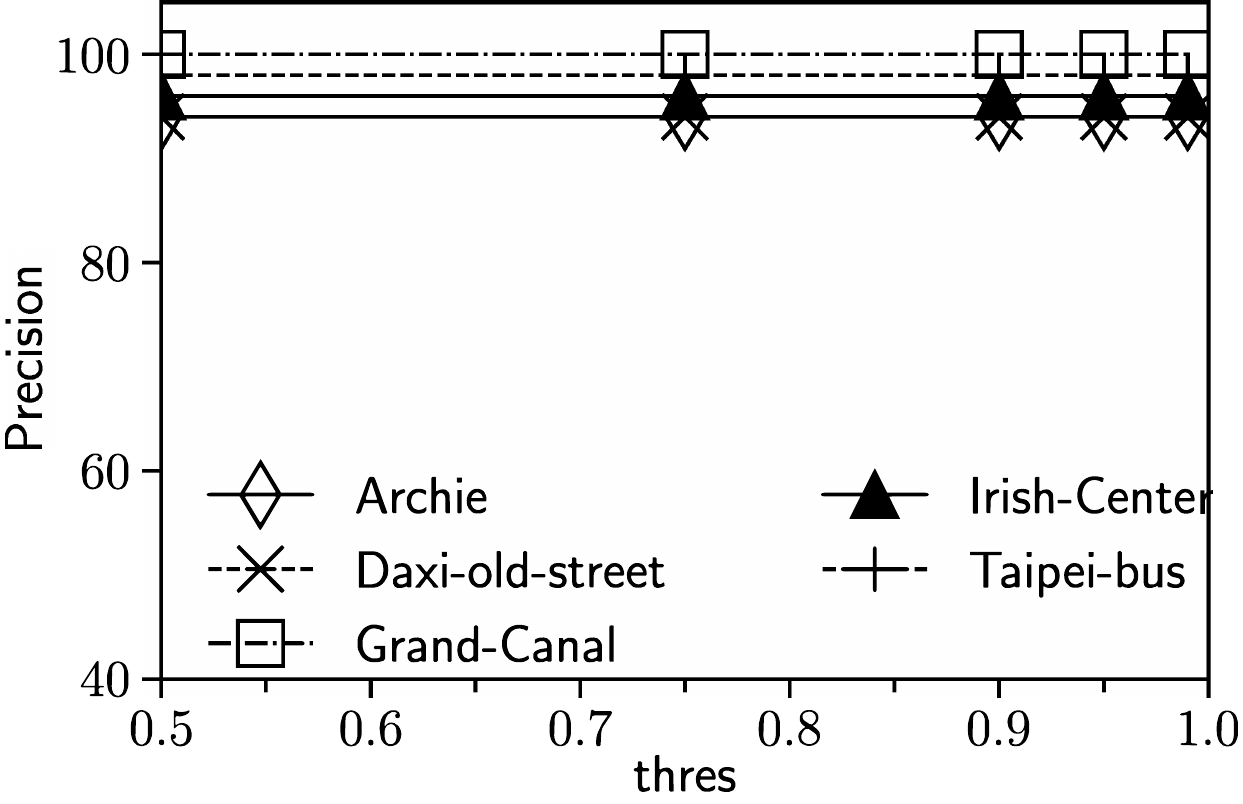}
    \caption{Precision}
    \label{fig:my_label}
    \end{subfigure}
    \hfill
    \begin{subfigure}{0.245\linewidth}
    \centering
    \includegraphics[width=\linewidth]{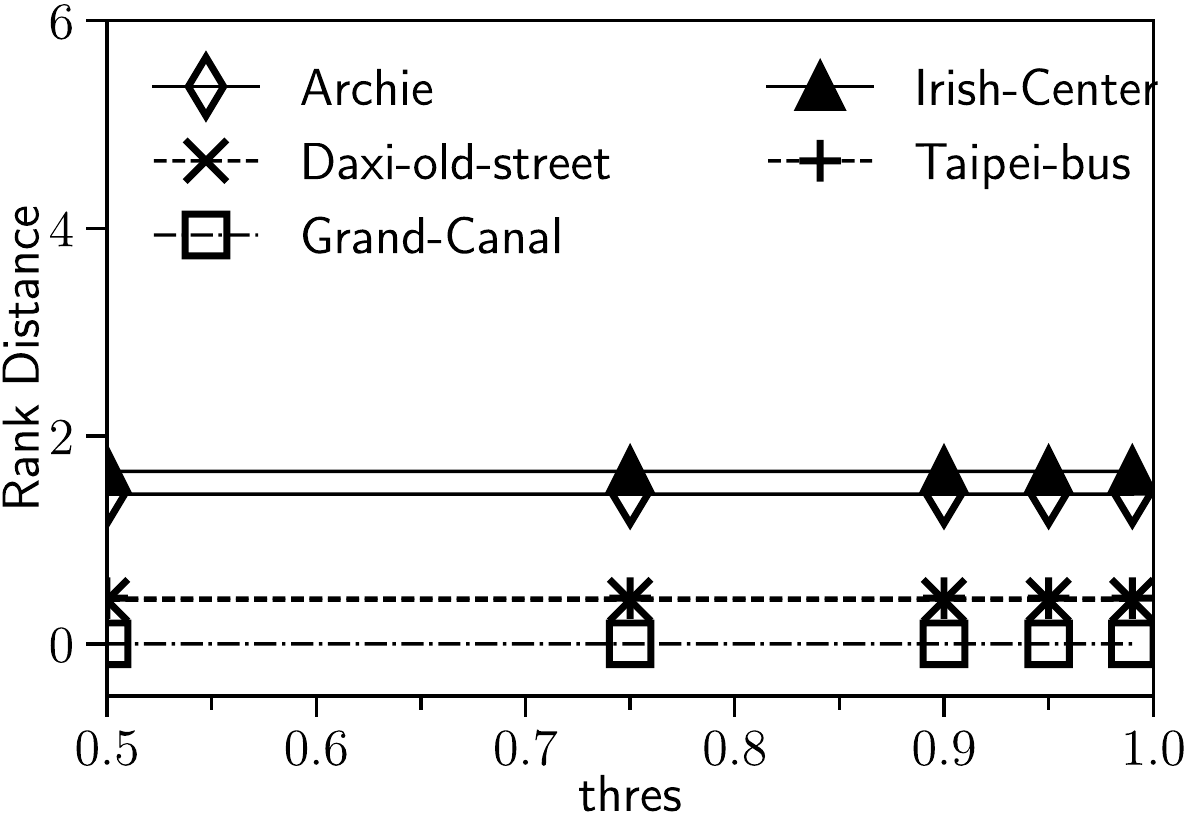}
    \caption{Rank distance}
    \label{fig:my_label}
    \end{subfigure}
    \hfill
    \begin{subfigure}{0.245\linewidth}
    \centering
    \includegraphics[width=\linewidth]{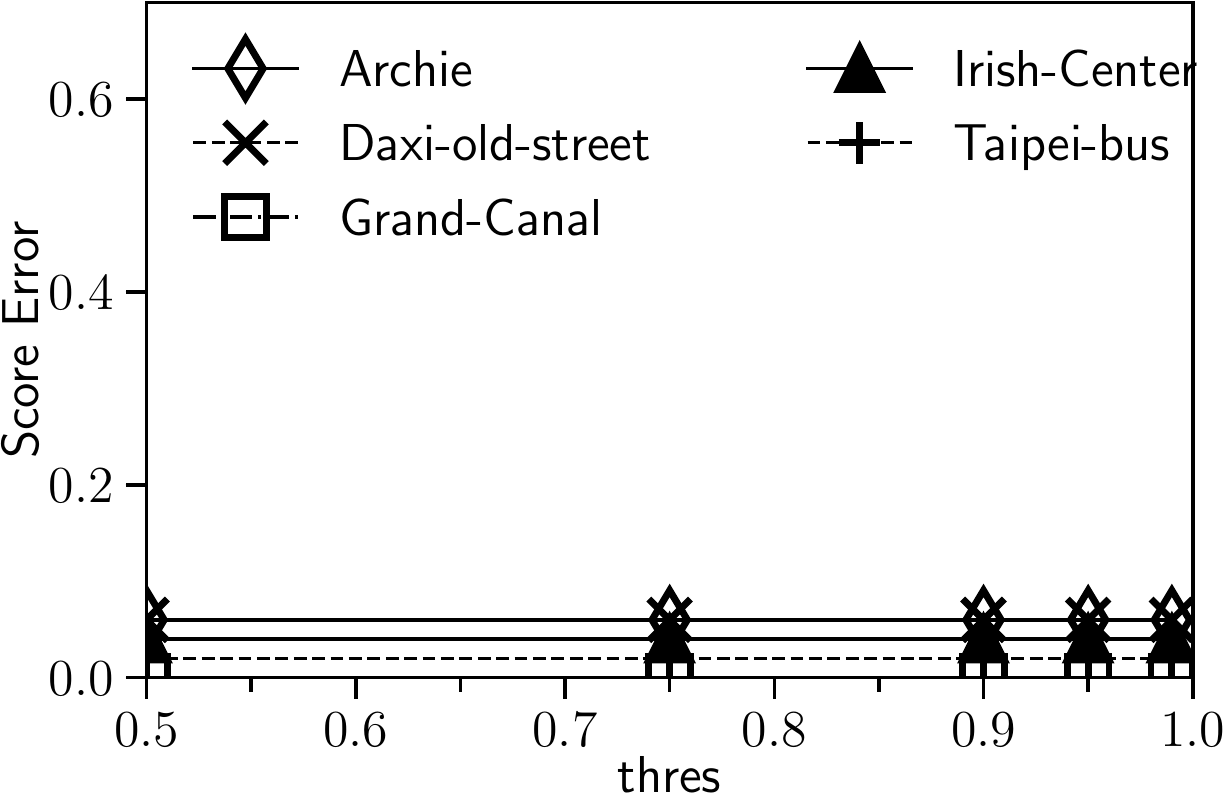}
    \caption{Score error}
    \end{subfigure}
    \caption{Impact of the confidence threshold $\thr$}
    \label{fig:3}
\end{figure*}

\vspace{-0.15cm}
\subsection{Overall Result and Comparisons}\label{sec:1}

Our first experiment aims 
to give an overall evaluation of \name.
We 
run the default Top-50 query (\thr=0.9) on five real videos.
Figure \ref{fig:1} shows the evaluation results.
The variation of the speedup on different datasets 
could be attributed to 
many factors
such as the video quality 
as well as the distributions of the object-of-interests.
Nonetheless, \name~is able to achieve a significant speedup of 16.3$\times$ to 18.3$\times$ over the scan-and-test baseline.
The result quality of \name~is also excellent.
Query precision values are all over 90\%, which are coherent with 
the 0.9 probability threshold of the queries.  
More specifically, 
all queries get results of very small rank distance, 
which indicates that \name~returns almost perfect result with an order of speedup over the exact baseline.
That can be further evidenced by observing that the score errors are all less than 0.1. 

\added{
\META{MR1}\RI{R1O2} \RII{R2O4}\RIII{R4O5}
Concerning the baselines,
surprisingly, 
HOG not only has zero to close-to-zero precision all the time,
but also runs slower than \name.  
The poor precision of HOG is expected
because its score errors are much higher than the oracle (scan-and-test)
and score errors between frames would lead to large errors 
in their relative rankings.
Although HOG is not based on deep learning, 
it is slow because it requires hundreds of SVM invocations
on many sub-regions per image.
For CMDN-only, 
the experimental results indicate that CMDN could serve
as the first phase of \name, 
but not as a standalone system by itself.
That is because running CMDN alone yields only a maximum of 1.6\% precision among all datasets.
TinyYOLOv3-only cannot be regarded as a true competitor 
neither.  With so few layers used, 
its precision and score error are no better than HOG.

Apart from the oracle scan-and-test,
select-and-topk is the only baseline that has over 0.9 precision.  
It has good precision because we manually tuned its $\lambda$ on each dataset.  
However, that precision comes with a huge cost
-- the experimental results 
show that select-and-topk is as slow as  scan-and-test even though we have given it all the advantages.
After we carefully look into their implementation, we found 
that those selection-only systems perform well on \emph{point query}
(e.g., finding frames that have ``cars''),
but not on \emph{range query} (e.g., finding frames that have \emph{more than} 10 cars).
That result justifies the need to develop 
specific techniques for Top-K video analytics.

Overall, \name~is the only winner in
both efficiency and accuracy.
Since the non-deep method HOG
and light-NN methods TinyYOLO-only and CMDN-only
all yield zero to near-zero precision,
we no longer include them in our further experiments below.
Furthermore, since the performance of select-and-topk is similar to the oracle scan-and-test 
but the former requires our manual tuning and measuring advantages, we retain only the scan-and-test method
as the baseline for measuring speed-up in the subsequent experiments.
}

\subsection{Zooming into \name} 
Table \ref{tab:breakdown} shows a breakdown of the end-to-end query runtime of \name. 
Most execution time ($\ge$80\%) is 
spent on Phase 1 to 
populate the initial uncertain table
because of the high volumes of frames processed by the CMDN.
The cost of Phase 1 gets relatively smaller for longer videos because we cap the sample size as 30000 frames.
In fact, Phase 1
can be done offline   
during data ingestion
(e.g., Focus \cite{focus})
or even at the edge where the videos are produced \cite{videoedge}.
But we make no such assumption here in this paper.

Phase 2 spent most of the time 
confirming the frames using the \replaced{GTOD}{oracle}.
Nonetheless,
that is almost minimal
because the fraction of
frames being cleaned 
is very small (0.16\%--0.76\%).
The algorithmic overheads are also minimal.
In fact, the fractions of time spent executing the functions {\sf Top-K($D_i^c$)} and {\sf Topk-prob}($D_i$, $\rhat_i$) are both less than 0.01\% and thus we do not show them in the table.
That indicates that our algorithmic optimizations are very effective.

\deleted{
From Figure \ref{fig:1}, 
we see that the select-and-topk baseline yields almost no speedup over the scan-and-test baseline even we have given it all the advantages
(e.g., not counting its training time).
That is because we found that these selection-only systems
perform well on \emph{point query} 
(e.g., finding frames that have ``cars'')
but poor on \emph{range query} (e.g., finding frames that have \emph{more than} 10 cars).
The result quality of the select-and-topk baseline is not shown in Figure \ref{fig:1}
because we manually tune their $\lambda$ to ensure precisions over 0.9.  Thus, we exclude them for clarity. 
Since the select-and-topk baseline does not offer any notable performance gain over the scan-and-test baseline and is impractical (requires manual tuning of $\lambda$), we drop it from our following discussion that evaluates \name~in detail.
}

\subsubsection{\bf Impact of K}\label{sec:2}
To understand the impact of K on \name,   
we run Top-K queries using different K values:
5, 10, 25, 50, 75, and 100.

Figure \ref{fig:2} shows that \name~has consistently high speedup in different values of K. 
Generally, \name~offers slightly better speedup when K is small.
That is because 
a smaller K results in a smaller result set, 
and thus the ``threshold'' frame tends to have a higher score (i.e., a higher $S_{\tf}$).
That in turns implies a higher $\phat_i$ based on Equation \ref{eq:topk_prob} 
and so that it can stop earlier by reaching {\tt thres} easier.

While the accuracy remains high for different values of K,
we observe that small K values tend to influence the precision more.
That is natural because the 
precision is a fraction influenced by the result size.
For example, the precision of \name~drops below 90\% 
on {\tt Irish-center} even though \name~missed only two frames out of the Top-10 result, resulting in a 80\% precision.  
Nonetheless, we observe that the result quality is actually high
from the other two quality metrics when K is small.
Therefore, if we look at the result accuracy 
using all three quality metrics, 
we can see that \name~produces high-quality results across different values of K.

\subsubsection{\bf Impact of \thr}\label{sec:3}

To understand the impact of the probability threshold on \name, we run Top-50 queries using different \thr~values:
0.5, 0.75, 0.9, 0.95, and 0.99.

Figure \ref{fig:3} shows that 
the value of \thr~is not crucial as long as it is over 0.5.
This is expected because we mentioned that $\phat_i$ 
actually improves
exponentially with the number of frames cleaned 
according to Equation \ref{eq:topk_prob}. 
In our experiments, it took 99\% of iterations to reach a probability threshold of 0.5 but only 1\% of iterations to reach 0.99.
This is indeed a nice result
because it confirms that \name~can 
hide this parameter from users in real deployments.

\begin{figure*}[h]
    \begin{subfigure}{0.245\linewidth}
    \centering
    \includegraphics[width=\linewidth]{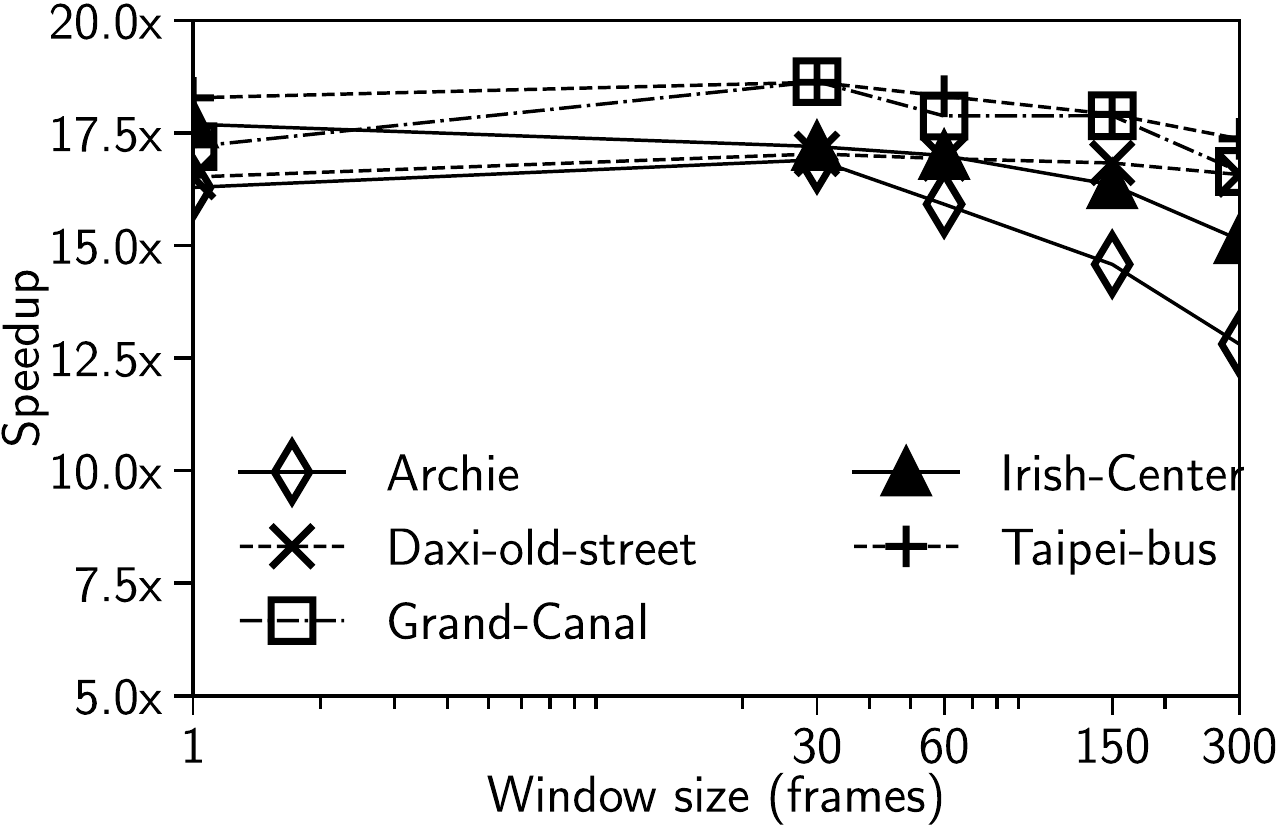}
    \caption{Speedup}
    \label{fig:my_label}
    \end{subfigure}
    \hfill
    \begin{subfigure}{0.245\linewidth}
    \centering
    \includegraphics[width=\linewidth]{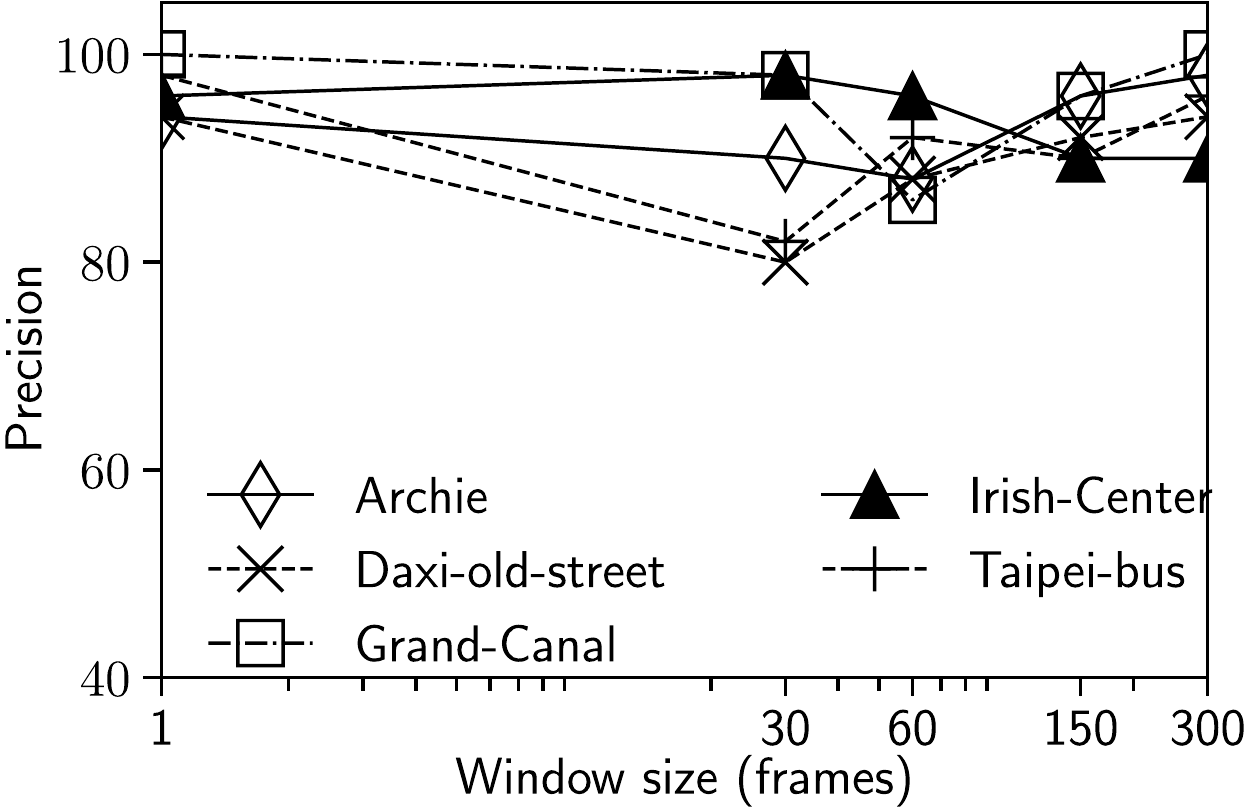}
    \caption{Precision}
    \label{fig:my_label}
    \end{subfigure}
    \hfill
    \begin{subfigure}{0.245\linewidth}
    \centering
    \includegraphics[width=\linewidth]{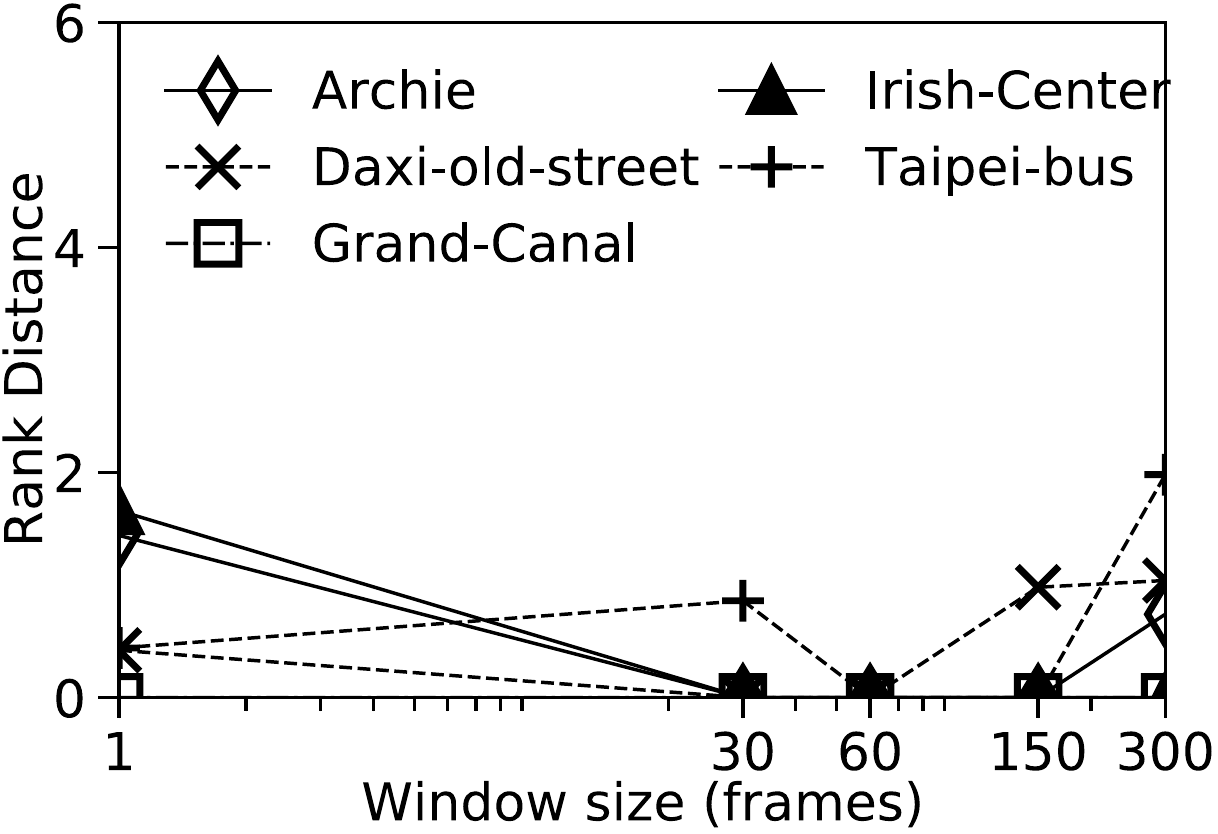}
    \caption{Rank distance}
    \label{fig:my_label}
    \end{subfigure}
    \hfill
    \begin{subfigure}{0.245\linewidth}
    \centering
    \includegraphics[width=\linewidth]{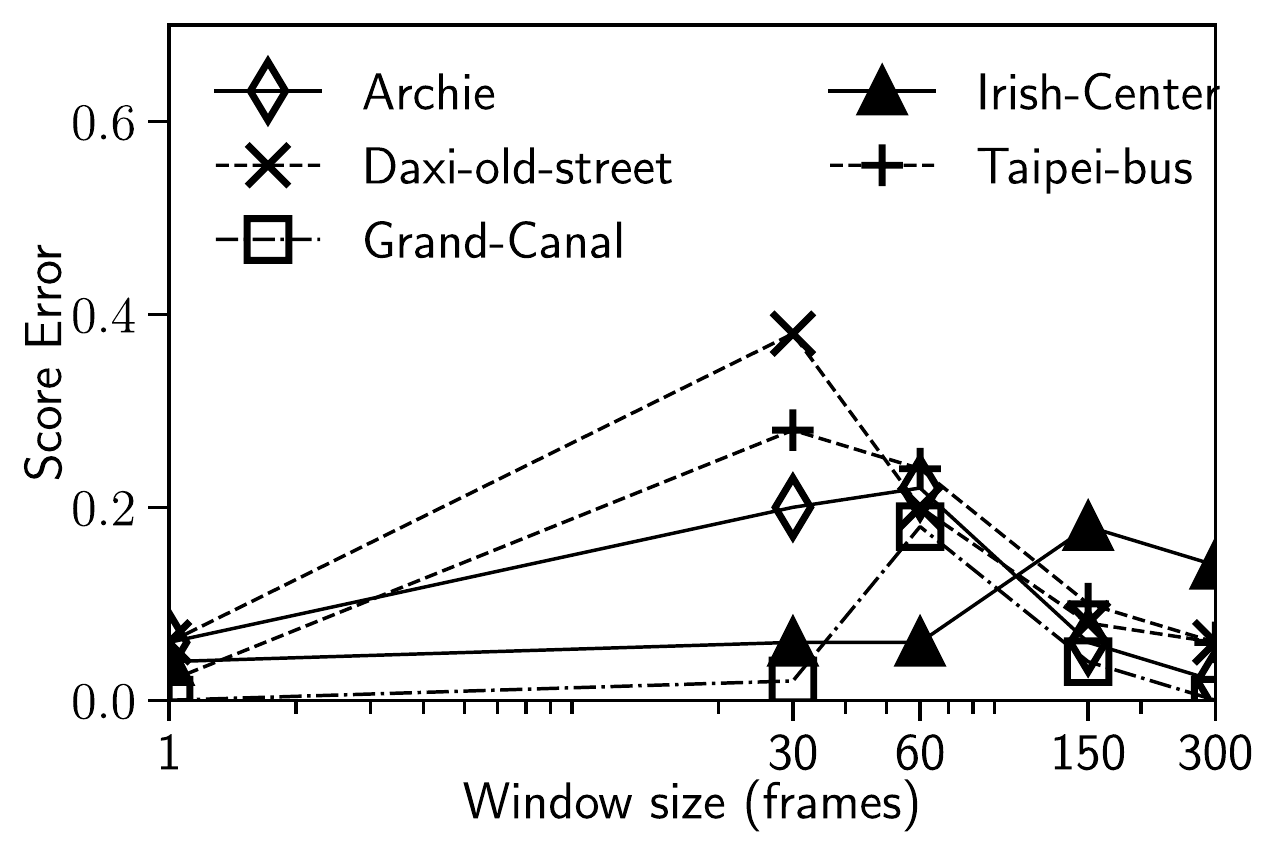}
    \caption{Score error}
    \end{subfigure}
    \caption{Varying the window size}
    \label{fig:5}
\end{figure*}

\begin{figure*}[h]
    \begin{subfigure}{0.245\linewidth}
    \centering
    \includegraphics[width=\linewidth]{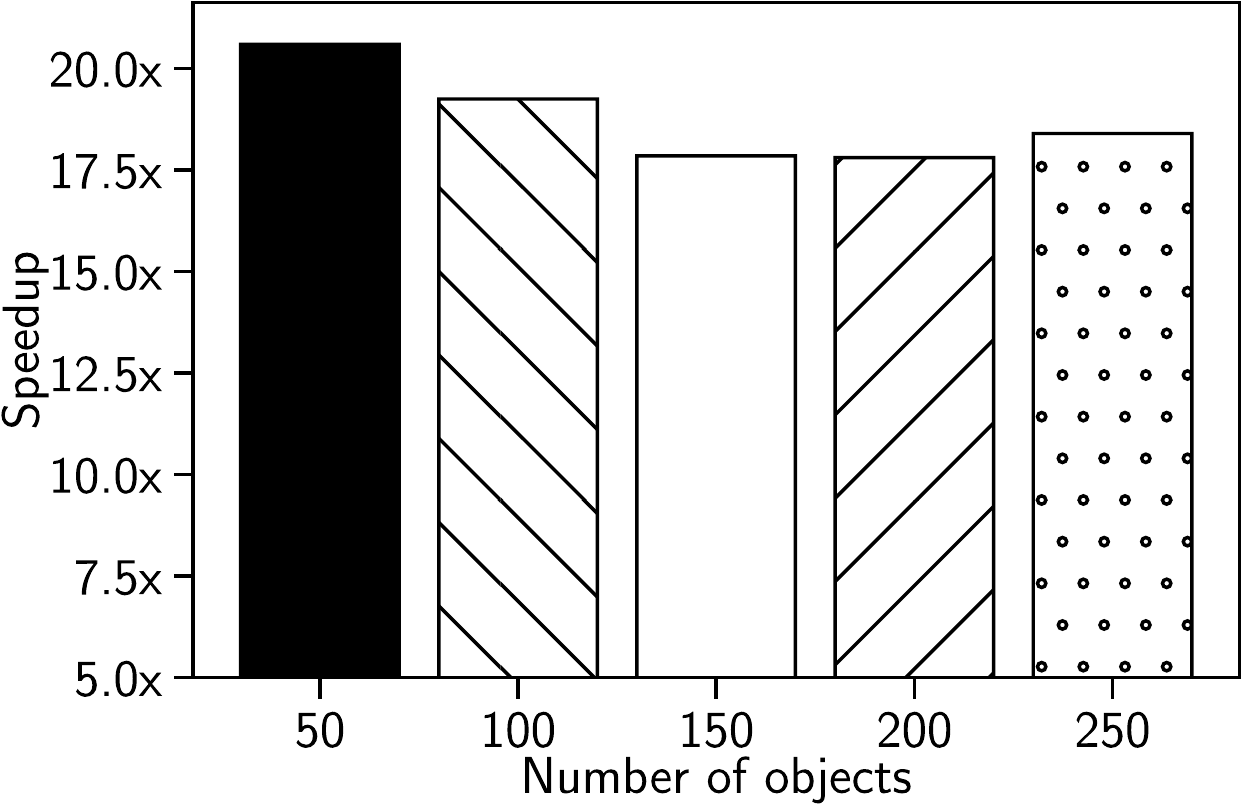}
    \caption{Speedup}
    \label{fig:my_label}
    \end{subfigure}
    \hfill
    \begin{subfigure}{0.245\linewidth}
    \centering
    \includegraphics[width=\linewidth]{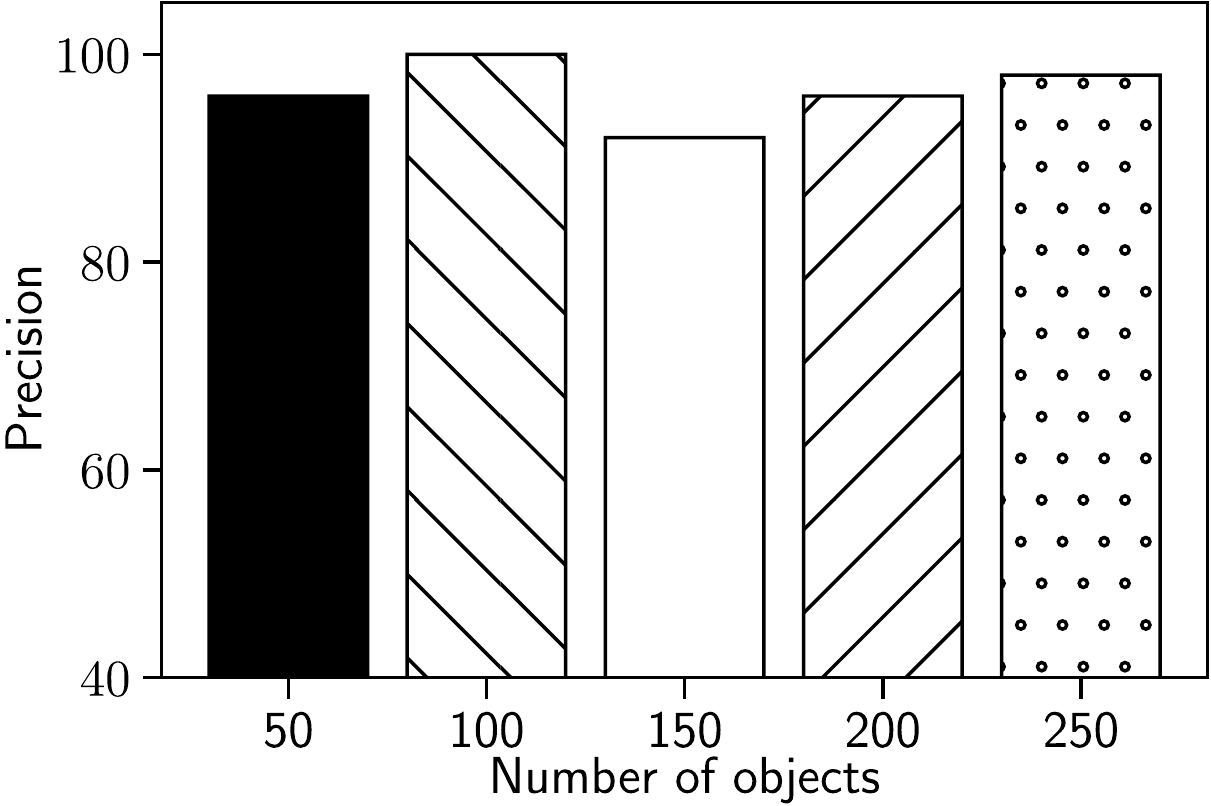}
    \caption{Precision}
    \label{fig:my_label}
    \end{subfigure}
    \hfill
    \begin{subfigure}{0.245\linewidth}
    \centering
    \includegraphics[width=\linewidth]{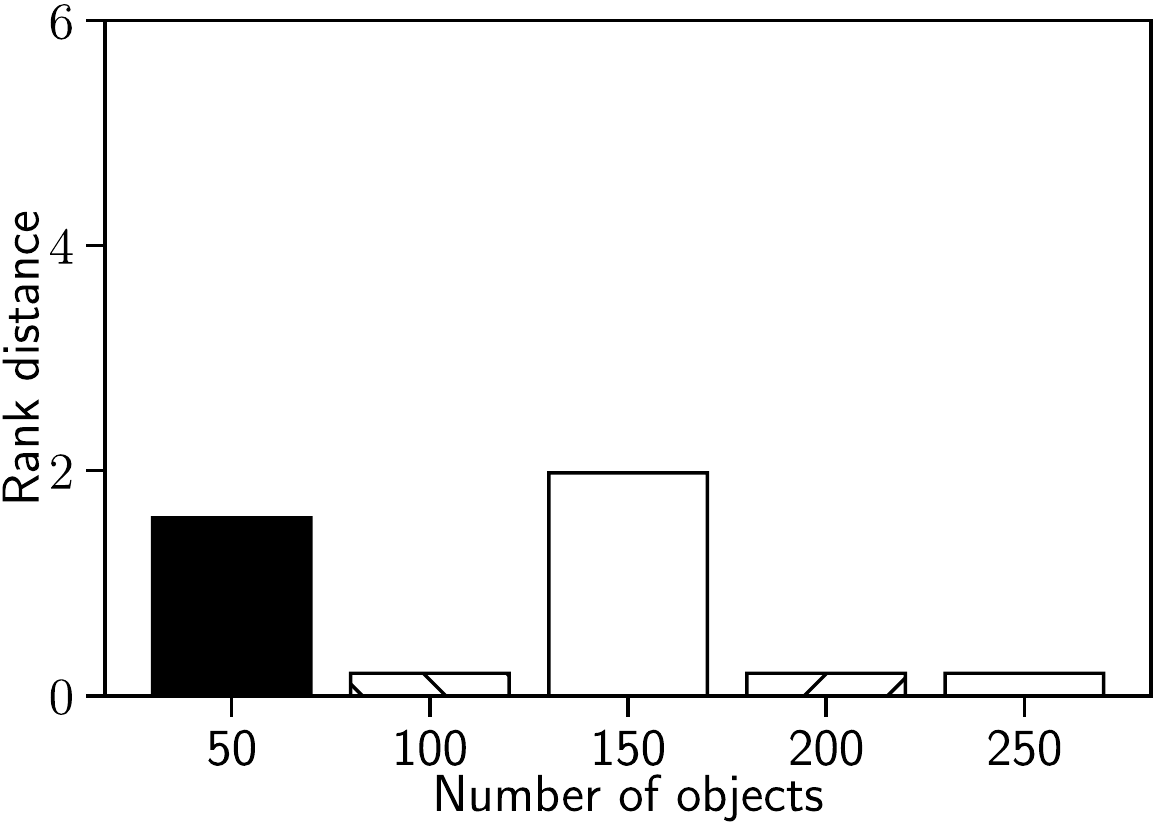}
    \caption{Rank distance}
    \label{fig:my_label}
    \end{subfigure}
    \hfill
    \begin{subfigure}{0.245\linewidth}
    \centering
    \includegraphics[width=\linewidth]{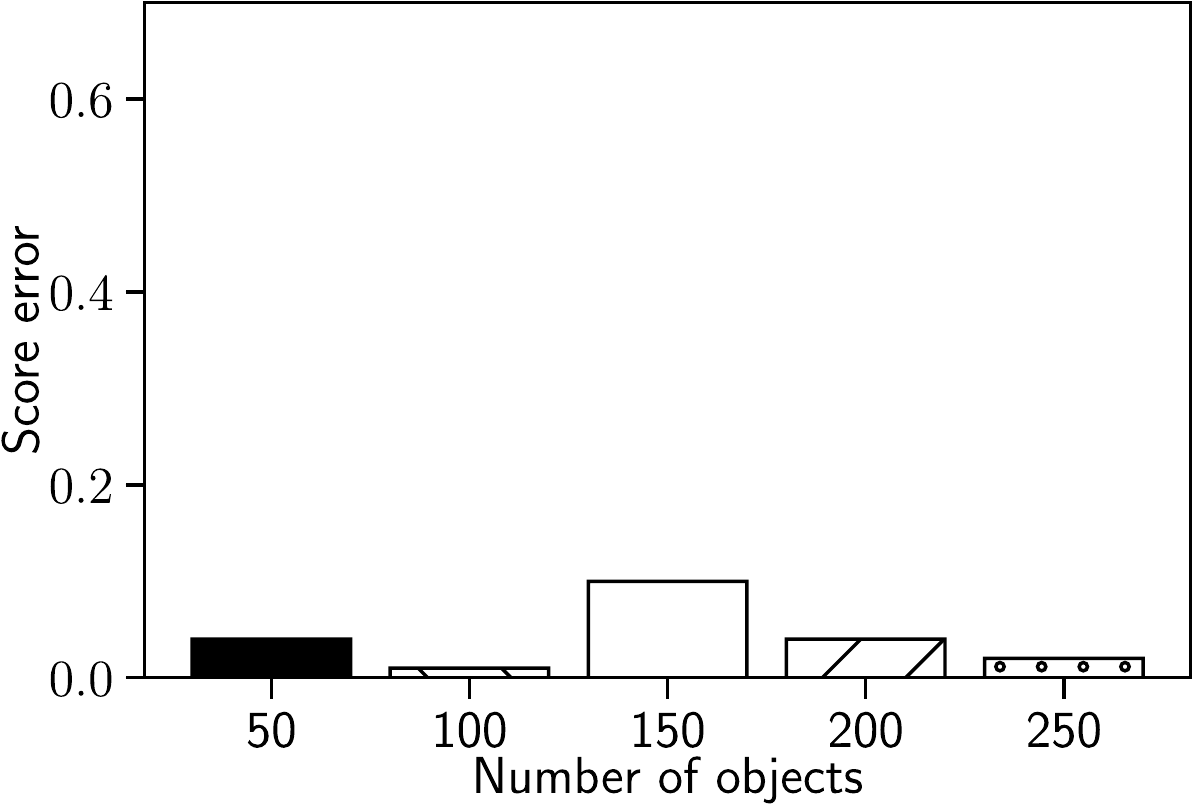}
    \caption{Score error}
    \end{subfigure}
    \caption{Varying the number of objects in Visual Road}
    \label{fig:6}
\end{figure*}
\begin{figure*}
    \begin{subfigure}{0.245\linewidth}
    \centering
    \includegraphics[width=\linewidth]{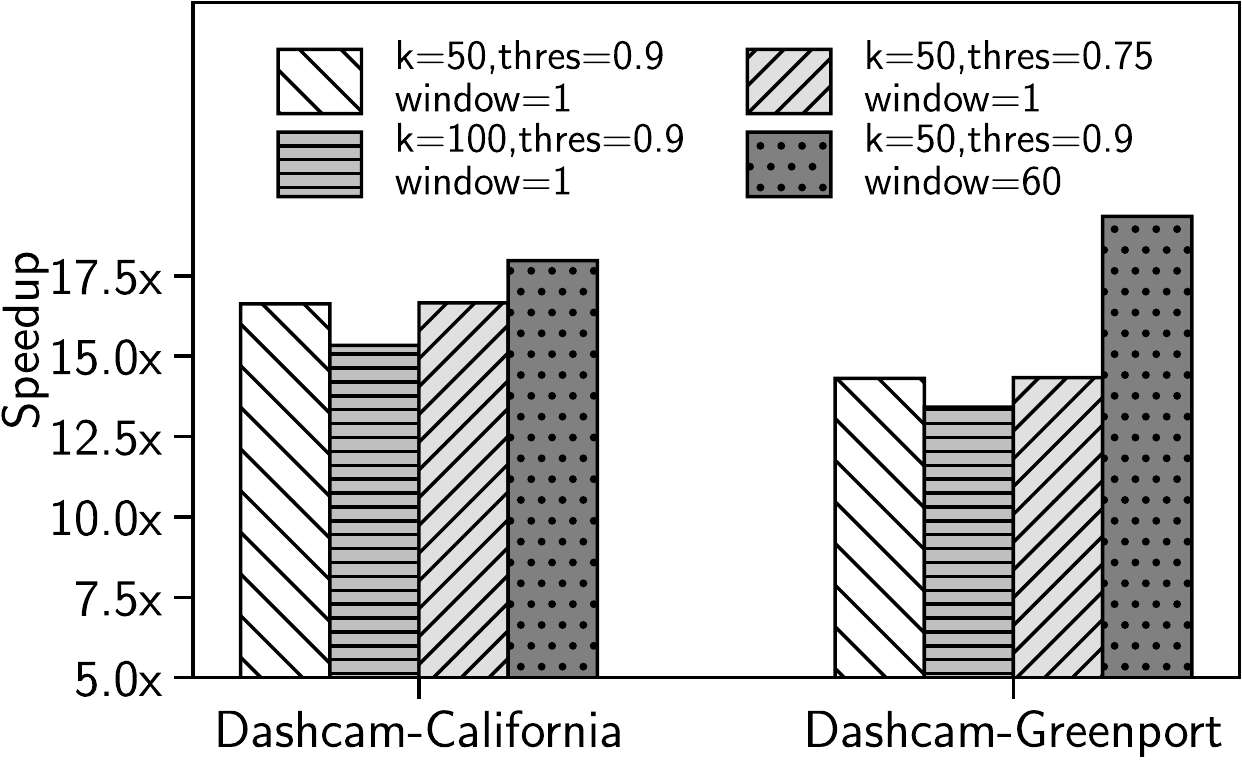}
    \caption{Speedup}
    \label{fig:my_label}
    \end{subfigure}
    \hfill
    \begin{subfigure}{0.245\linewidth}
    \centering
    \includegraphics[width=\linewidth]{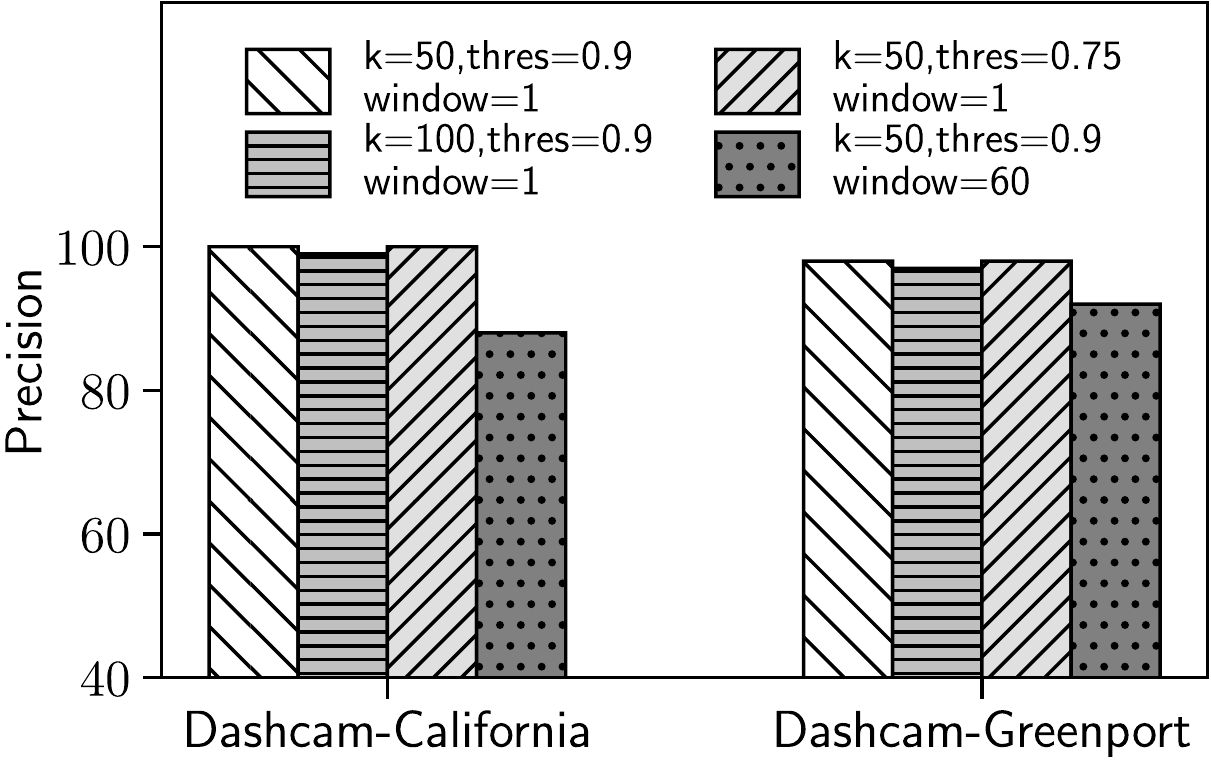}
    \caption{Precision}
    \label{fig:my_label}
    \end{subfigure}
    \hfill
    \begin{subfigure}{0.245\linewidth}
    \centering
    \includegraphics[width=\linewidth]{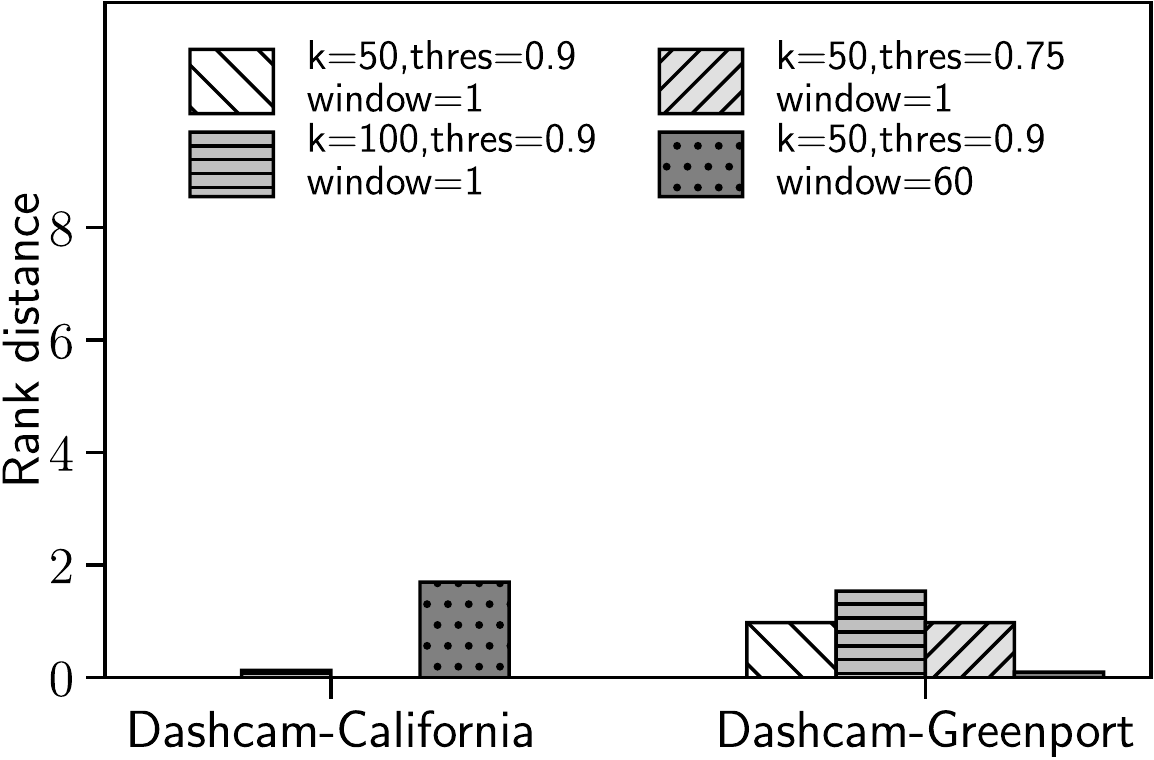}
    \caption{Rank distance}
    \label{fig:my_label}
    \end{subfigure}
    \hfill
    \begin{subfigure}{0.245\linewidth}
    \centering
    \includegraphics[width=\linewidth]{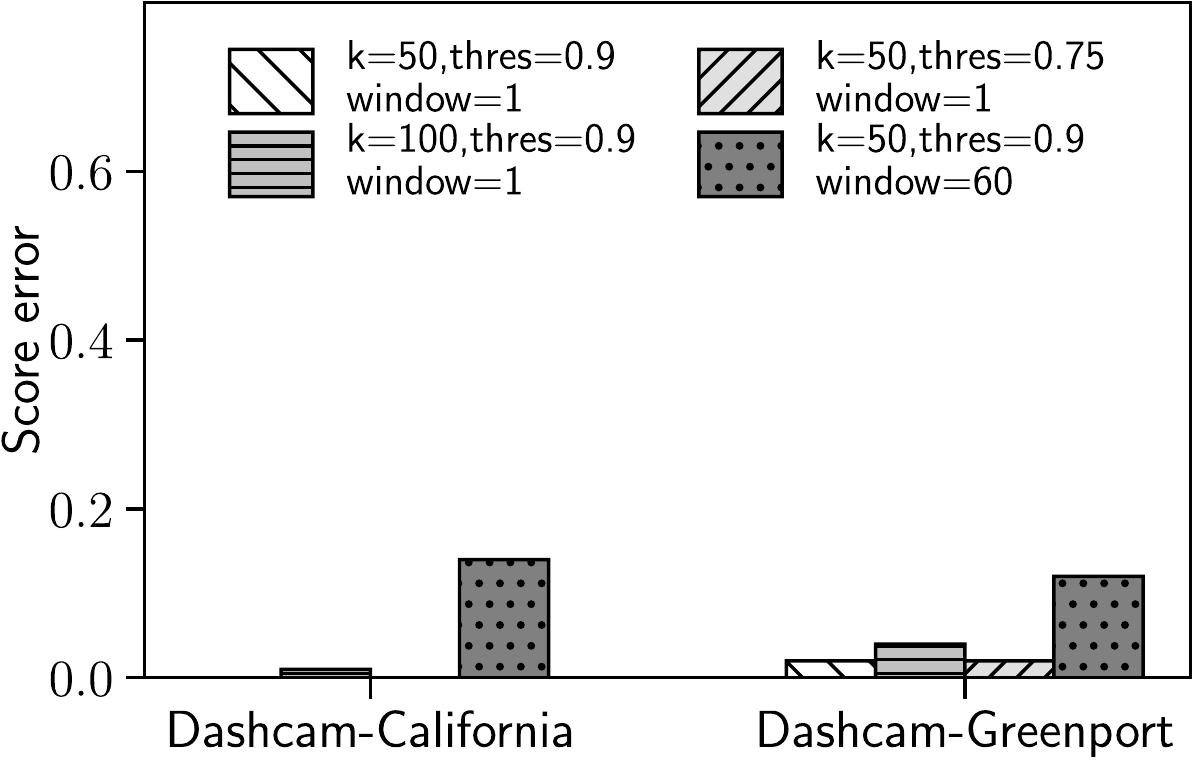}
    \caption{Score error}
    \end{subfigure}
    \caption{\added{Scoring function using a deep depth estimator}}
    \begin{varwidth}[t]{1cm}
    \RICAP{R1O1}
    \end{varwidth}
    \label{fig:4}
\end{figure*}

\subsubsection{\bf Top-K Windows}\label{sec:5}

To 
evaluate the 
window feature of \name,
we run Top-50 window queries 
with different window sizes: no window (i.e., a window of 1 frame), 30, 60, 150, and 300 frames.
The probability threshold is still 0.9.  
In Phase 2, 
each window samples 10\% of its frames to infer the ground-truth.

Figure \ref{fig:5} shows that \name~performs 
as good as frame-based Top-K,
but the speedup drops slightly when the window size gets larger. There are two reasons.
First, a larger window size essentially reduces the number of windows.
As shown in the experiments above, our Top-K algorithm is smart at picking the most promising frame out of millions of frames to clean.
A reduced number of windows would, however, reduce the number of choices \name~has. 
Second, a larger window implies 
more frames have to be confirmed by the \replaced{GTOD}{oracle} per selected candidate (compared with only one frame has to be confirmed per selected candidate when no window is specified).

In terms of accuracy, 
the result quality remains high in general.
Sometimes, the precision may fluctuate a bit 
because of the randomness in sampling.
For example, the precision of {\tt Taipei-bus} drops slightly below 0.9 when the window size is 30 frames. 
The fluctuation diminishes when the window size gets larger because of the larger sample size.

\subsubsection{\bf Impact of Object Density}\label{sec:6}
To evaluate \name's default object counting UDF in greater detail,
we generate five 30-fps videos using the Visual Road benchmark \cite{visualroad}
because we cannot control that in real videos.
Each generated video is 
 ten hours long
 in 416$\times$416 resolution.
All five synthetic videos share 
the same setting except the 
total number of cars.
Specifically, 
the videos are all taken
by the same ``camera'' shooting from the same ``angle''
of the same ``mini-city''.
The only moving objects in the videos are cars
and we control 
the total number of cars in that mini-city
from 50 to 250. 
During the data generation process, we uncovered a problem in the Visual Road benchmark 
in which we could only stably generate at most 15 minutes long video.  
After discussing with the authors of Visual Road, 
we concatenated 40 clips of 15-minute videos to form each ten-hour video.

Figure \ref{fig:6} shows the evaluation result on Visual Road under the default Top-50 query. 
We observe a speedup of 17.8$\times$ to 20.6$\times$
with excellent accuracy.
This experiment indicates that the good performance of \name~would not be affected much by the content of the videos.

\subsubsection{\bf Scoring Function}\label{sec:7}
\added{
In this last experiment, \RII{R1O1}
we use \name~to answer another type of query using another UDF.
Specifically, we build a UDF that uses 
the deep depth-estimator in \cite{depth_estimation} 
and pose Top-K queries 
over two dashcam videos we found 
on Youtube (the last two rows in Table \ref{tab:data}).
This experiment is to recall the fleet management use case mentioned in the introduction. The objective is to find the most dangerous tailgating moments.

Figure \ref{fig:4} shows the results
on various scenarios.
These include
the default Top-50 (\thr=0.9) query,
one Top-{\bf 100} (\thr=0.9) query,
one Top-50 query with a smaller threshold ({\bf \thr=0.75}),
and one Top-50 (\thr=0.9) {\bf window} query.
We observe that \name~maintains its
high result quality ($>$90\% precision)
with about 15$\times$ speedup over the baseline scan-and-test approach under all those different settings.
}

\deleted{
\input{related_work}
}
\section{Conclusions and Future Work} \label{sec:conclusion}
With a massive amount of video data available and 
generated incessantly, the discovery of interesting information from videos becomes an exciting area of data analytics. 
Although deep neural networks enable semantic extraction from videos with human-level accuracy, they are unable to process video data
at scale unless efficient systems are built on top.
State-of-the-art video analytics systems have not supported rich analytics like Top-K.  In response, we build \name, a fast and accurate Top-K video analytics system. 
To our knowledge, 
\name~is the first video analytics system 
that treats the uncertain output of deep networks
as a first-class citizen 
and provides probabilistic guaranteed accuracy.  
Currently, 
\name~is a standalone system that supports only ranking. 
\added{
Richer analytics can be enabled by integrating it with 
an expressive video query language or libraries like FrameQL \cite{blazeit} and Rekall \cite{rekall}. \RIII{R4O1}
Finding the skyline \cite{skyline} from such uncertain video data is also an interesting future work.}  \RII{R2O3}
\name~is an open-source project. 
It is available at: \texttt{https://github.com/everest-project/everest}.

\begin{acks}
This work is partly supported by Hong Kong General Research Fund (14200817), Hong Kong AoE/P-404/18, Innovation and Technology Fund (ITS/310/18, ITP/047/19LP) and  Centre for Perceptual and Interactive Intelligence (CPII) Limited under the Innovation and Technology Fund.
\end{acks}

\clearpage
\bibliographystyle{ACM-Reference-Format}
\bibliography{cite}  %

\end{document}